%% file: main_arxiv.tex
\documentclass[nonacm]{acmart}

\AtBeginDocument{%
  }

\setcopyright{none}
\acmDOI{}
\acmISBN{}
\acmConference{}{}{}

\definecolor{benchmarkBasic}{HTML}{0F9DCC}
\definecolor{benchmarkEasy}{HTML}{2E9B49}
\definecolor{benchmarkMedium}{HTML}{D79B00}
\definecolor{benchmarkHard}{HTML}{C96A1B}
\definecolor{benchmarkExpert}{HTML}{C63D3D}

\usepackage{xcolor}
\usepackage{tcolorbox}
\usepackage{colortbl}
\usepackage{listings}
\usepackage{tikz}
\usepackage{tabularx}
\usepackage{multirow}
\usetikzlibrary{arrows.meta,positioning,fit,backgrounds}
\usepackage{graphicx}
\usepackage{subcaption}
\usepackage[ruled,vlined]{algorithm2e}

\lstdefinelanguage{PCBGPTPython}[]{Python}{
  morekeywords={},
}

\lstdefinestyle{pcbgpt}{
  language=PCBGPTPython,
  basicstyle=\ttfamily\footnotesize,
  keywordstyle=\bfseries\color{blue!65!black},
  stringstyle=\color{red!55!black},
  commentstyle=\itshape\color{green!45!black},
  showstringspaces=false,
  breaklines=true,
  breakatwhitespace=true,
  columns=fullflexible,
  keepspaces=true,
  numbers=none,
  frame=none,
  xleftmargin=0pt,
  framexleftmargin=0pt,
}

\newcommand{\red}[1]{\textcolor{red}{#1}}
\renewcommand{\red}[1]{#1}
\newcommand{\systemName}{pcbGPT}

\input{tables/automatic_results_variables.tex}

\input{tables/human_comparator_agreement_variables.tex}
\input{tables/failure_taxonomy_variables.tex}
\input{tables/failure_correlation_variables.tex}

\begin{document}

\title{\systemName: Automatic PCB Schematic Synthesis from Natural Language Requirements}

\author{Tobias King}
\affiliation{%
  \institution{Karlsruhe Institute of Technology}
  \city{Karlsruhe}
  \state{Baden-Wuerttemberg}
  \country{Germany}
}
\email{tobias.king@kit.edu}

\author{Steven Kehrberg}
\affiliation{%
  \institution{Bosch Sensortec GmbH}
  \city{Reutlingen}
  \state{Baden-Wuerttemberg}
  \country{Germany}
}
\email{steven.kehrberg@bosch-sensortec.com}

\author{Michael Beigl}
\affiliation{%
  \institution{Karlsruhe Institute of Technology}
  \city{Karlsruhe}
  \state{Baden-Wuerttemberg}
  \country{Germany}
}
\email{michael.beigl@kit.edu}

\author{Tobias Röddiger}
\affiliation{%
  \institution{Karlsruhe Institute of Technology}
  \city{Karlsruhe}
  \state{Baden-Wuerttemberg}
  \country{Germany}
}
\email{tobias.roeddiger@kit.edu}

\renewcommand{\shortauthors}{King et al.}

\begin{abstract}
  Translating natural-language hardware requirements into correct printed circuit board (PCB) schematics remains difficult in embedded, IoT, and wearable development. Designers must choose compatible components, interpret datasheets, add support circuitry, and expose correct interfaces before layout and prototyping can begin, while many such circuits cannot be validated through straightforward simulation. We present \systemName, a grounded system for generating editable KiCad schematics from natural-language specifications. \systemName{} represents circuits in a Python DSL and combines tool-augmented synthesis with component-library search, datasheet-grounded design knowledge, execution-based checking, structural and semantic validation, and an interactive web workflow that supports iterative refinement and synchronization with KiCad projects. We evaluate the system on 20 embedded schematic-generation tasks with reference implementations, required components, and interface constraints that enable automatic comparison. The best model reaches overall pass@1 of \red{\AutomaticResultsBestOverallPassAtOne{}} and pass@5 of \red{\AutomaticResultsBestOverallPassAtFive{}}; pass@1 is \red{\AutomaticResultsBestBasicPassAtOne{}} on basic and easy tasks, \red{\AutomaticResultsBestMediumPassAtOne{}} on medium tasks, and \red{\AutomaticResultsBestHardPassAtOne{}} on hard tasks. These results, together with failure analysis, show that \systemName{} can already generate useful, reviewable first-draft schematics for early prototyping, but is not yet reliable enough to replace expert review.
\end{abstract}


\begin{teaserfigure}
  \centering
  \includegraphics[page=15, width=\textwidth]{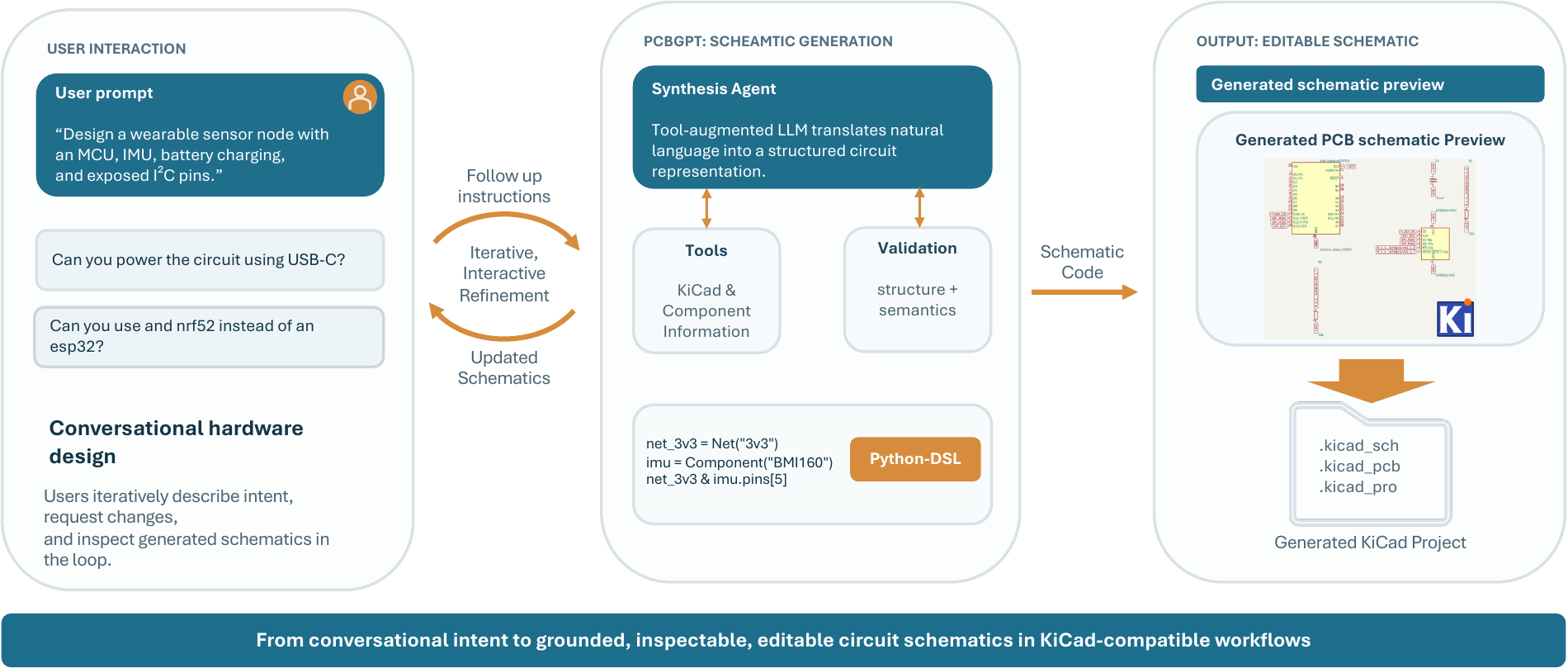}
  \centering
  \caption{\systemName{} transforms conversational natural-language hardware requirements into editable KiCad-compatible schematic drafts through grounded component selection, structured circuit synthesis, deterministic checking, and semantic validation.}
  \Description{Three-panel overview of \systemName{}. The left panel shows conversational hardware design: a user provides a natural-language prompt asking for a wearable sensor node with a microcontroller, IMU, battery charging, and exposed I2C pins, then follows up with requests to use USB-C power and to display the schematic. The center panel shows the grounded synthesis pipeline, where pcbGPT uses component grounding, a circuit DSL with components, pins, and nets, deterministic checking, and semantic validation to translate the request into a structured intermediate circuit representation. The right panel shows the resulting editable schematic draft, with example nets such as SDA, SCL, 3V3, and GND connecting an MCU and IMU in a KiCad-compatible schematic. Arrows indicate the progression from user interaction to grounded synthesis to an inspectable draft schematic.}
  \label{fig:teaser}
\end{teaserfigure}

\maketitle
\input{chapters/introduction.tex}
\input{chapters/background.tex}
\input{chapters/methology.tex}
\input{chapters/benchmark_design.tex}

\input{chapters/experiments.tex}

\input{chapters/failure_taxonomy.tex}

\input{chapters/discussion.tex}

\begin{acks}
This work was funded by the German Federal Ministry of Education and Research (BMBF) within the Software Campus 3.0 program under grant number 01IS23066.
\end{acks}

\bibliographystyle{ACM-Reference-Format}
\bibliography{zotero, literature}

\appendix
\input{appendix/benchmark_inventory.tex}

\end{document}

%% file: tables/automatic_results_variables.tex
\providecommand{\AutomaticResultsBestModelName}{gpt-5.3-codex}
\providecommand{\AutomaticResultsBestOverallPassAtOne}{0.90}

\providecommand{\AutomaticResultsBestOverallPassAtFive}{1.00}
\providecommand{\AutomaticResultsBestBasicPassAtOne}{1.00}

\providecommand{\AutomaticResultsBestMediumPassAtOne}{0.91}

\providecommand{\AutomaticResultsBestMediumPassAtFive}{1.00}
\providecommand{\AutomaticResultsBestHardPassAtOne}{0.72}

\providecommand{\AutomaticResultsBestHardPassAtFive}{1.00}

\providecommand{\AutomaticResultsBestHardAverageSimilarity}{1.00}
\providecommand{\AutomaticResultsSecondModelName}{qwen3.5-397b-a17b}
\providecommand{\AutomaticResultsSecondOverallPassAtOne}{0.72}

\providecommand{\AutomaticResultsSecondOverallPassAtFive}{0.90}
\providecommand{\AutomaticResultsSecondBasicPassAtOne}{1.00}

\providecommand{\AutomaticResultsSecondMediumPassAtOne}{0.71}

\providecommand{\AutomaticResultsSecondHardPassAtOne}{0.28}

%% file: tables/human_comparator_agreement_variables.tex
\providecommand{\HumanComparatorEvaluatedRuns}{400}

\providecommand{\HumanComparatorAgreementPct}{97.5}
\providecommand{\HumanComparatorKappa}{0.941}

\providecommand{\HumanComparatorFalseNegative}{10}

%% file: tables/failure_taxonomy_variables.tex
\providecommand{\FailureTaxonomyEvaluatedRuns}{400}
\providecommand{\FailureTaxonomyFailedRuns}{115}
\providecommand{\FailureTaxonomyFailureRatePct}{28.8}
\providecommand{\FailureTaxonomySupportingComponentTypeCount}{68}
\providecommand{\FailureTaxonomySupportingComponentTypeSharePct}{59.1}
\providecommand{\FailureTaxonomySupportingComponentValueCount}{39}
\providecommand{\FailureTaxonomySupportingComponentValueSharePct}{33.9}
\providecommand{\FailureTaxonomyConfigurationCount}{26}
\providecommand{\FailureTaxonomyConfigurationSharePct}{22.6}
\providecommand{\FailureTaxonomyInterfaceCount}{46}
\providecommand{\FailureTaxonomyInterfaceSharePct}{40.0}
\providecommand{\FailureTaxonomyTopologyCount}{31}
\providecommand{\FailureTaxonomyTopologySharePct}{27.0}

\providecommand{\FailureTaxonomyModelGptFiveThreeCodexFailureRatePct}{8.0}

\providecommand{\FailureTaxonomyModelQwenThreeFiveLargeFailureRatePct}{28.0}

\providecommand{\FailureTaxonomyModelGptFiveOneFailureRatePct}{35.0}

\providecommand{\FailureTaxonomyModelQwenThreeFiveSmallFailureRatePct}{44.0}

%% file: tables/failure_correlation_variables.tex
\providecommand{\FailureCorrelationEvaluatedRuns}{400}

\providecommand{\FailureCorrelationPromptCount}{20}
\providecommand{\FailureCorrelationTokenPointBiserialR}{0.481}
\providecommand{\FailureCorrelationWorkingMedianTokens}{134{,}997}
\providecommand{\FailureCorrelationNonWorkingMedianTokens}{355{,}143}
\providecommand{\FailureCorrelationComponentPearsonR}{0.863}
\providecommand{\FailureCorrelationComponentSpearmanRho}{0.883}

%% file: chapters/introduction.tex
\section{Introduction}

Designing electronic circuits for ubiquitous, embedded, IoT, and wearable systems remains a technically demanding task.
Devices such as physiological sensing nodes, environmental monitors, connected sensing platforms, or activity trackers \cite{roddigerOpenEarable20OpenSource2025,tangTRingSmartRing2025,lepoldHARNodeTimeSynchronisedOpenSource2026}, require the careful selection of compatible components, correct interconnections, and adherence to numerous implicit design conventions.
While modern electronic design automation (EDA) tools such as KiCad \cite{kicad} provide powerful support for schematic design and printed circuit board (PCB) layout, they assume substantial prior expertise in electronics and hardware design.
Users must understand datasheets, voltage compatibility, communication protocols such as I\textsuperscript{2}C or SPI, pull-up resistor requirements, decoupling strategies, and common reference design patterns.
As a result, hardware prototyping in ubiquitous computing, IoT, and wearable system development continues to require either domain specialists, significant iteration cycles or both \cite{roddigerOpenEarable20OpenSource2025,lepoldHARNodeTimeSynchronisedOpenSource2026,tangTRingSmartRing2025}.
This barrier slows down rapid experimentation and limits participation of interdisciplinary researchers, designers, and practitioners who may understand the intended device behavior but lack formal training in electrical engineering.
We therefore position \systemName{} as an interdisciplinary prototyping aid for early schematic exploration, while the current workflow remains expert-facing at the review stage: generated schematics are intended to be inspected, refined, and accepted by hardware-literate users rather than used as unchecked final designs.
Therefore, in this work, we investigate how agentic AI can aid users during the schematic-generation step of the printed circuit board (PCB) design process.

Recent advances in large language models (LLMs) have demonstrated strong capabilities in structured reasoning, code generation, and tool use \cite{chenEvaluatingLargeLanguage2021}.
LLM-based systems can translate natural language descriptions into executable programs \cite{chenEvaluatingLargeLanguage2021} and iteratively refine outputs using feedback \cite{madaan2023selfrefineiterativerefinementselffeedback}.
These capabilities suggest the possibility of leveraging LLMs to automate aspects of hardware design, such as schematic generation from high-level specifications in natural language.
However, applying LLMs to hardware design, particularly schematic generation for embedded, IoT, and wearable systems, poses distinct challenges that differ fundamentally from software, HDL synthesis, or other analog hardware design tasks.
Compared with software and VHDL, embedded circuit design presents a less favorable setting for LLM-based generation.
Software artifacts are abundantly represented in training corpora, and many models are explicitly tuned for programming tasks, such as OpenAI Codex \cite{chenEvaluatingLargeLanguage2021} or Anthropic's models used with Claude Code \cite{ClaudeCodeAnthropic}.
Moreover, software, HDL and some analog designs can typically be executed or simulated, enabling precise feedback for validation and iterative refinement \cite{laiAnalogCoderAnalogCircuit2025,shenAtelierAutomatedAnalog2024}.
By contrast, mixed-signal circuits composed of microcontrollers, sensors, power supplies, and peripheral interfaces often lack simulation models, making correctness harder to verify and increasing reliance on component-level electrical constraints and design experience.
Even slight errors in generated circuits can lead to non-functional designs, which are often only identified during physical prototyping, resulting in costly iteration cycles.

In this work, we present \systemName, an LLM-driven system for automated printed circuit board (PCB) schematic generation from natural language specifications.
Our focus is the early design stage, in which a designer uses datasheets and reference circuits to select components, define interfaces, and specify electrical connections at the schematic level, rather than performing downstream PCB placement, routing, and manufacturing preparation.
When processing user requirements into functional circuit designs, \systemName{} automatically selects suitable components from the locally installed KiCad library, retrieves relevant design information from datasheets using a custom knowledge extraction approach, and performs structural and semantic validation of the generated design before outputting an editable KiCad schematic draft.
The system further supports both fully automatic generation and human-in-the-loop interaction through a web-based interface that allows users to iteratively refine requirements, inspect intermediate circuit representations, and synchronize generated projects with KiCad workflows.
Inputs to the system can be either natural-language descriptions of desired functionality or existing KiCad projects that the system can extend or modify based on user instructions.
The output of the system is an editable KiCad project draft containing the generated schematic, which can be opened, inspected, modified, and used as a starting point for downstream EDA workflows.

Our approach consists of three building blocks: a synthesis agent, a set of component tools, and a validation stage.
The synthesis agent interprets the user requirements and proposes candidate circuit designs by orchestrating the component tools, which ground this process in the local KiCad environment by allowing the synthesis agent to search for suitable symbols, retrieve datasheets, and extract component-specific design knowledge needed for correct integration.
Instead of relying on electrical simulation which is often infeasible for complex circuits, we introduce a validation stage that first executes the generated design and performs deterministic structural checks before applying semantic review to assess whether the circuit plausibly satisfies the design intent.
Errors identified during validation are returned to the synthesis agent, forming an iterative generate, execute, and repair loop that refines the design until it passes all checks. 

We evaluate our system with a series of reference circuits and expert review.
We introduce a set of circuit design problems with solutions which are representative of ubiquitous, IoT, and wearable scenarios.
These tasks specify required, key components and interfaces where applicable, enabling automated comparison between generated circuits and intended solutions; we additionally assess generated circuits from this set with domain experts.
Through this evaluation, we assess the technical correctness and practical limitations of the system.

In summary, our contributions include:
\begin{itemize}
    \item \textbf{\systemName}: A grounded system for natural-language-driven automatic schematic generation that produces editable KiCad project drafts compatible with established EDA workflows.
    \item A Python-based circuit DSL for schematic synthesis and evaluation, supporting structured circuit construction, optional components, flexible pin assignments, and export to KiCad-compatible artifacts.
    \item A tool-augmented LLM pipeline that combines grounded component selection, datasheet-informed synthesis, and iterative validation for embedded circuit design.
    \item A web-based interaction workflow that supports iterative prompting, schematic inspection, and synchronization with KiCad projects, extending the automatic generator toward practical KiCad-centered use.
    \item An empirical evaluation of schematic generation on 20 representative embedded, IoT, and wearable circuit design tasks, using reference implementations and interface constraints for automated comparison alongside expert review of generated circuits.
\end{itemize}

%% file: chapters/background.tex
\section{Related Work}
We organise related work into three areas most relevant to our system: LLM-based program synthesis and textual circuit representations, grounding and validation for real PCB components, and interactive hardware prototyping systems.

\subsection{LLM-Driven Circuit Synthesis and Code-Based Representations}
Large language models have demonstrated strong capabilities in program synthesis, structured reasoning, and API orchestration \cite{chenEvaluatingLargeLanguage2021}.
Code-oriented evaluation frameworks such as HumanEval established the paradigm of translating natural language into executable programs that are then assessed via functional correctness \cite{chenEvaluatingLargeLanguage2021}.
Agentic prompting strategies such as ReAct \cite{yaoReActSynergizingReasoning2023a}, tool-calling approaches \cite{schickToolformerLanguageModels2023}, and program-aided formulations \cite{gao2023palprogramaidedlanguagemodels} show how external tools and deterministic runtimes can improve correctness on structured generation tasks.
Iterative refinement methods, including self-debugging and self-critique pipelines, further motivate architectures in which one component generates an artefact and another evaluates or repairs it \cite{madaan2023selfrefineiterativerefinementselffeedback,chen2023teachinglargelanguagemodels}.
Because LLMs operate primarily over text, it is natural to cast circuit generation in a textual form as well, using a representation the model can produce reliably and that downstream tools can deterministically execute.

This motivates the use of programmatic circuit design as the intermediate layer in our pipeline, since board-level hardware description languages and code-first PCB workflows already provide textual representations that compile into conventional design artefacts.
Polymorphic Blocks introduces a board-level HDL supporting multiple abstraction layers and export to conventional layout tools \cite{linPolymorphicBlocksUnifying2020}.
Weaving Schematics and Code presents an interactive IDE synchronising HDL code and visual schematic representations \cite{linWeavingSchematicsCode2021}.
Frameworks such as SkiDL \cite{SKiDLSKiDL}, Atopile \cite{IntroductionAtopile}, and PySpice \cite{1OverviewPySpice} demonstrate the benefits of code as an intermediate representation for abstraction, reuse, automation, and, where possible, simulation.

Related work in AI-assisted EDA similarly frames hardware design as a generation or optimisation problem.
Recent systems translate natural language into HDL or analogue circuit descriptions \cite{changChipGPTHowFar2023,blockloveChipChatChallengesOpportunities2023,laiAnalogCoderAnalogCircuit2025,shenAtelierAutomatedAnalog2024}, while other learning-based approaches target analogue sizing, topology search, fanout, or routing \cite{wangLearningDesignCircuits2020,vijayaraghavanAUTOCIRCUITRLReinforcementLearningDriven2025,liFanoutNetNeuralizedPCB2023}.
However, these systems are typically evaluated with simulation-based or formal design-rule oracles and are aimed at HDL design or narrower EDA subproblems rather than end-to-end PCB schematic generation.
Our system differs by treating board-level schematic generation as a tool-grounded program synthesis task in a circuit DSL that compiles directly into editable KiCad artefacts.
This circuit DSL is directly tailored to the needs of AI-assisted PCB design, with expressive error handling, grounding in real components, and a structure that supports downstream validation without requiring full simulation.

\subsection{Grounded Generation and Validation for Real Components}
Automated circuit generation requires more than producing structurally plausible code: the resulting design must correspond to real, library-available components with correct pin assignments and compatible interfaces.
Prior work emphasises grounding design decisions in datasheets, component libraries, and structured knowledge representations \cite{loAutoFritzAutocompletePrototyping2019,rameshTurningCodersMakers2017a}.
Recent systems use canonical pinout retrieval, structured component databases, and datasheet-derived knowledge to improve electrical plausibility and reduce hallucinated parts or invalid wiring \cite{SKiDLSKiDL}.
Datasheet-grounded verification and structural constraint checking have likewise been used to detect inconsistencies between intended behaviour and realised connectivity \cite{rameshTurningCodersMakers2017a,garza2025typedschematicsblockbasedpcbdesign}.

Validation remains particularly challenging for embedded and wearable systems because many important components, including microcontrollers, digital sensors, and communication modules, are not readily simulatable at the schematic level.
Traditional electrical rule checking and design rule checking detect common violations but do not guarantee functional correctness \cite{kicad}.
Simulation-based validation works well for analogue blocks or HDL-level designs when accurate models exist \cite{1OverviewPySpice}, but this assumption often breaks down for real PCB assemblies.
This has motivated alternatives such as structural rules, interface typing, and semantic compatibility checks; TypedSchematics, for example, introduces type-like circuit interfaces to detect invalid compositions \cite{garza2025typedschematicsblockbasedpcbdesign}.
The remaining gap in related work is a pipeline that couples real-component grounding with validation strategies tailored to non-simulatable embedded hardware.
Our system addresses this by integrating datasheet retrieval and KiCad library search into synthesis and by using a separate validation agent that performs structural and semantic checks on the generated schematic rather than relying on full simulation.

\subsection{Interactive Hardware Prototyping Systems}
Human-computer interaction and ubiquitous computing research has long aimed to lower the barriers to hardware prototyping, especially for non-expert users.
Trigger-Action Circuits synthesises circuitry from high-level behavioural specifications and supports interactive exploration of design alternatives \cite{andersonTriggerActionCircuitsLeveragingGenerative2017}.
AutoFritz provides autocomplete-style assistance in virtual breadboarding environments grounded in datasheet schematics and community circuit corpora \cite{loAutoFritzAutocompletePrototyping2019}.
Embedded Design Generation formulates device synthesis as a constraint-solving problem from high-level functional descriptions \cite{rameshTurningCodersMakers2017a}.
More recent ubiquitous computing systems such as MorphingCircuit \cite{wangMorphingCircuitIntegratedDesign2020}, Wearable Bits \cite{jonesWearableBitsScaffolding2020}, SkinLink \cite{kuSkinLinkOnbodyConstruction2023}, and CircuitGlue \cite{lambrichtsCircuitGlueSoftwareConfigurable2023} extend this space across self-morphing, wearable, on-body, and modular electronics prototyping.
These systems emphasize interactive, hardware-grounded prototyping, but not interactive KiCad project generation from natural-language requirements.

%% file: chapters/methology.tex
\section{Methodology}

\subsection{System Overview}
The objective of \systemName{} is to translate a natural-language circuit specification into a KiCad-compatible schematic that can be inspected, edited, and reused in a standard PCB design workflow.
We treat the problem as a code generation task which we solve using three system components: a \emph{synthesis agent} that plans and assembles the design, a set of \emph{component tools} that ground the agent in the local KiCad libraries and component datasheets, and a \emph{validation stage} that executes, checks, and critiques candidate designs.
This decomposition reflects the engineering steps required for early-stage schematic development: first identifying the needed functional blocks, then consulting concrete component information, and finally checking whether the resulting design is structurally coherent and reviewable.

The synthesis agent is responsible for converting the user prompt into an executable schematic description.
It does not rely purely on parametric memory, but instead issues tool calls to identify suitable KiCad symbols, recover schematic-relevant component knowledge, and iteratively revise the design based on execution and validation feedback.
The component tools provide the grounding required for this process by connecting the language model to the exact component names, pin maps, footprint candidates, and datasheet-derived integration rules available in the local design environment.
The validation stage then checks whether the generated design is syntactically executable, compatible with the DSL and KiCad environment, and structurally and semantically plausible with respect to the original design intent.
Together, these three parts form an iterative generate, execute, repair loop.
The synthesis agent proposes a circuit, the validation stage returns concrete failure signals or semantic critiques, and the agent updates the design until a valid solution is obtained or a predefined iteration limit is reached.
\autoref{fig:system-overview} provides a graphical overview of this pipeline.

\input{figures/system_overview}

\subsection{Synthesis Agent}
\label{sec:agents}

At the top level, schematic generation is driven by a tool-augmented synthesis agent.
Its input is a natural-language request such as a regulator stage, sensor interface, or complete embedded subsystem, and its output is a circuit expressed in our Python-based schematic representation.
The agent must therefore bridge the gap between informal design intent and the exact symbol, pin, and connectivity requirements of the KiCad ecosystem.

In contrast to a plain language-model baseline, the synthesis agent does not attempt to recall all component details from memory.
Instead, it follows a staged workflow.
First, it identifies the functional blocks implied by the prompt and selects concrete candidate components.
Second, for unfamiliar or complex parts, it retrieves component-specific design knowledge needed for correct integration.
Third, it assembles the circuit and submits it to the execution and validation pipeline.
If the candidate design fails due to invalid symbols, incorrect footprints, illegal pin references, missing support circuitry, or functional inconsistencies, the returned feedback is used to revise the next candidate.

This generate, execute, repair loop is central to the overall methodology.
It converts tool output and validation feedback into a corrective learning signal inside the same synthesis episode.
As a result, the model is not forced to solve the task in a single pass, but can progressively converge toward a design that is both executable in the target environment and consistent with component-specific wiring constraints.

\subsection{Component Tools}
\label{sec:tooling}

The synthesis process is grounded by tools that expose the information sources an engineer would normally consult during schematic capture.
These tools connect the agent to the local KiCad installation and to component documentation, allowing it to reason over exact symbol identities and implementation constraints instead of relying on approximate recall.

\paragraph{Component search tool.}
The component search tool allows the agent to retrieve relevant parts from the local KiCad library collection using natural-language queries such as ``buck regulator,'' ``USB-C connector,'' or ``resistor.''
For each candidate, the tool returns the library name, symbol name, pin descriptors, and candidate footprint options.
This grounding step is necessary because the final circuit must reference valid KiCad symbols and pins exactly as they exist in the local environment.
Internally, the tool combines embedding-based retrieval with lexical search.
The agent prompt requires queries to target component classes rather than highly specific parameter strings (e.g. Search for "Resistor" instead of "4.7k resistor"), which encourages symbol reuse and reduces redundant search calls.

\paragraph{Component information tool.}
Selecting the correct symbol is not sufficient for schematic synthesis; the agent must also understand how a component should be integrated into a circuit.
For this purpose, the component information tool builds a reusable, design-oriented summary for a specific KiCad part identified by its library and symbol name.
The summary contains the information most relevant for schematic construction, including pin behavior, required and recommended external components, typical application circuits, operating constraints, sequencing requirements, and other wiring rules needed for correct use.

Internally, the tool follows a multi-stage pipeline.
It first resolves the component from the local KiCad library and retrieves the associated datasheet; if the component-provided URL fails, the implementation falls back to PDF web-search to obtain the needed datasheet.
The tool renders every datasheet page to an image at 300 DPI, extracts page text, and groups pages into batches of up to 45 pages per request.
For each batch, the model receives the page images together with the extracted text and is instructed to produce a page-local schematic-design summary covering pin behavior, required support circuitry, typical application circuits, values, and usage constraints.
The batch summaries are then merged in a second step that deduplicates repeated information and keeps the most specific version when chunks overlap.
To avoid repeated computation, the final summary can be cached on disk.
When a datasheet-model request times out, the implementation retries once after a short delay; if extraction still fails, the component is reported as unavailable to the calling agent instead of silently inventing information, which allows the synthesis agent to fall back to alternative components.

These component tools define the grounded knowledge boundary of the synthesis system.
The search tool answers which concrete part can be used in the design, while the component information tool answers how that part should be wired and supported in practice.
Together they provide the external context required before circuit assembly and validation can begin.

\subsection{Validation}
\label{sec:validation-agent}

Once the synthesis agent produces a candidate circuit, the design enters the validation stage.
This stage combines deterministic execution checks with a validation agent that critiques the resulting schematic at a higher semantic level.
Its role is not only to confirm that the code can be executed, but to determine whether the resulting circuit is structurally coherent and plausibly satisfies the original prompt.

\paragraph{Circuit execution.}
The first step is to execute the generated circuit code and instantiate the design in the local environment.
During this step, the system checks that the Python code is syntactically valid and that it satisfies the constraints of the circuit representation and the local KiCad installation.
Common failures at this stage include invalid component identifiers, missing or incompatible footprints, illegal pin references, and other violations of DSL constraints.
If instantiation succeeds, the system performs ERC to detect common connectivity and rule violations before exporting the design artifacts.

\paragraph{Validation agent.}
After a circuit passes execution and ERC, a dedicated validation agent evaluates the schematic for structural and semantic correctness.
The validator receives the generated circuit representation together with the user defined requirements.
For nontrivial components, it can also use the datasheet-derived summaries described above.
This allows it to detect problems that would not be visible from code execution alone, such as omitted support circuitry, incorrect interface exposure, incompatible subsystem wiring, or missing external connections required by the task.

If the validation agent identifies a problem, it returns a structured list of root-cause issues to the synthesis agent.
These issues are then used in the next repair round.
In this way, validation acts as a semantic feedback mechanism layered on top of the lower-level execution checks.
When all checks pass, the pipeline compiles the resulting design into a KiCad project draft and associated export artifacts for downstream use.

This reflects a methodological tradeoff: many embedded circuits cannot be validated through straightforward simulation, and a purely rule-based checker is too limited for heterogeneous real components. We therefore combine deterministic execution and ERC with probabilistic semantic validation, without treating these judgments as formal verification.

\autoref{tab:validation-responsibilities} summarizes the two validation stages. Deterministic execution and rule checking establish syntactic validity, while probabilistic semantic validation assesses whether the generated design is plausible with respect to component usage, support circuitry, and overall design intent. We treat validation as a design-assistance mechanism for early-stage schematic development rather than a substitute for expert review or physical verification.

\begin{table}[t]
\centering
\small
\caption{Summary of the two validation stages in \systemName{}.}
\label{tab:validation-responsibilities}
\begin{tabularx}{\columnwidth}{llX}
\toprule
\textbf{Stage} & \textbf{What is checked} & \textbf{Examples} \\
\midrule
\multirow{2}{*}{\begin{tabular}[t]{@{}l@{}}Syntactic validation\\(deterministic)\end{tabular}} & \begin{tabular}[t]{@{}l@{}}$\bullet$ Executability\\$\bullet$ Rule-based consistency in the DSL\\$\bullet$ KiCad environment compatibility\end{tabular} & Invalid DSL code, unknown symbols, invalid pin names, missing footprints, unconnected required pins, and shorted nets detected by ERC \\
& & \\
\midrule
\multirow{2}{*}{\begin{tabular}[t]{@{}l@{}}Semantic validation\\(probabilistic)\end{tabular}} & \begin{tabular}[t]{@{}l@{}}$\bullet$ Prompt-level plausibility\\$\bullet$ Correct component usage\\$\bullet$ Datasheet-derived integration knowledge\end{tabular} & Missing pull-ups or decoupling, swapped interface nets, incorrect voltage-domain integration, wrong configuration pins, or implausible local support circuitry \\
& & \\
\bottomrule
\end{tabularx}
\end{table}

\subsection{Code Representation for PCB Circuits}
PCB circuits are typically described as netlists, that is, textual descriptions of components and their interconnections.
However, netlists are not an ideal representation for LLM-based synthesis because they contain substantial redundant information and are under-represented in the training corpora of many state-of-the-art models.

To address this limitation, we developed a Python-embedded circuit DSL that serves as an intermediate representation between language-driven synthesis and KiCad export.
The DSL models an explicit object graph of \texttt{Circuit}, \texttt{Component}, \texttt{Net}, and \texttt{Pin} objects.
We chose Python as the host language because it is widely used and because large language models are comparatively strong at understanding and generating Python code.
The DSL integrates directly with the KiCad ecosystem, where components are organized into libraries and uniquely identified by their names.

The representation allows creating \textit{Circuit} objects to which \textit{Component} and \textit{Net} objects can be added.
Components wrap KiCad library symbols and expose their available pins, footprint options, and related attributes.
Pins can then be accessed by their identifiers, for example \texttt{r1.pins["1"]}.
If a referenced pin does not exist, the DSL raises informative errors that list the valid alternatives.
Connectivity is expressed through operator overloading.
Specifically, the bitwise-and operator \texttt{\&} is used to merge nets and attach pins, for example \texttt{net \& pin}.
Using \texttt{pin \& pin} merges the existing nets of both pins or creates an implicit net if neither pin is yet connected.

While existing systems such as SKiDL \cite{SKiDLSKiDL} also allow PCB circuits to be expressed as Python code, they do not provide several features required for our synthesis-and-evaluation setting, which motivated the development of our own AI-oriented DSL:

\begin{enumerate}
    \item \textbf{Pin error handling}: When an invalid pin is referenced, our DSL returns the list of valid pins for that component, enabling automated correction.
    \item \textbf{Value format}: Our DSL enforces normalized value strings for passives and forbids arbitrary \texttt{value} fields on non-passives, reducing LLM formatting errors; e.g., \texttt{"10k"} is valid for a resistor, but not for a microcontroller.
    \item \textbf{Footprint validation}: Our DSL rejects missing or invalid footprints immediately, enabling early recovery.
    \item \textbf{Optional components}: Our DSL can mark components as optional, distinguishing required structure from permissible extras during evaluation.
    \item \textbf{Flexible pin assignments}: Our DSL introduces \texttt{FlexiblePin}, which allows a single logical connection to map to a set of interchangeable physical pins defined in the reference design. This enables robust evaluation in cases where multiple pins can fulfill the same function (e.g., equivalent GPIOs on an MCU).
    \item \textbf{Evaluation-oriented export}: Our DSL supports export to JSON, graphs, netlists, and KiCad projects for both circuit capture and automated benchmark evaluation.
\end{enumerate}

\begin{figure}[t]
\centering
\begin{minipage}{0.95\columnwidth}
\lstset{style=pcbgpt}
\begin{lstlisting}
from src.Circuit import Circuit

circuit = Circuit()

r1 = circuit.add_component(name="R", library="Device", value="10k", footprint="Resistor_SMD:R_0603_1608Metric")
c1 = circuit.add_component(name="C", library="Device", value="100nF", footprint="Capacitor_SMD:C_0603_1608Metric")

vin  = circuit.add_net("VIN")
vout = circuit.add_net("VOUT")
gnd  = circuit.add_net("GND")

vin  & r1.pins["1"]
vout & r1.pins["2"] & c1.pins["1"]
gnd  & c1.pins["2"]
\end{lstlisting}
\end{minipage}
\caption{Example of the Python-embedded circuit DSL. The snippet constructs a minimal RC low-pass stage by instantiating KiCad resistor and capacitor components, creating named nets (VIN, VOUT, GND), and connecting pins with the overloaded \textit{\&} operator.}
\Description{Code example of the Python-embedded circuit DSL. It creates a circuit, adds one resistor and one capacitor with KiCad footprints, creates three nets named VIN, VOUT, and GND, and connects the resistor and capacitor pins using the overloaded ampersand operator to form a simple RC stage with an output capacitor to ground.}
\label{fig:circuit-dsl}
\end{figure}

\subsection{Interactive Web Workflow}
\label{sec:web-workflow}

In addition to fully automatic batch generation, \systemName{} supports an interactive browser-based workflow that turns generation into a persistent design session rather than a single-shot prompt-response exchange.
The web application maintains session history and synchronized project state across rounds, streams intermediate tool calls and results, and presents both generated circuit code and a rendered schematic preview.
Through this interface, users can iterate on requirements, inspect validation feedback, download the synthesized netlist or KiCad project archive, and continue after local KiCad edits.
The web interface acts as both a front-end to the generation agent and a coordination layer between conversational design intent, executable synthesis artifacts, and KiCad-based editing.
An overview of the web-based workflow is shown in figure \autoref{fig:web-workflow}.

\input{figures/web_workflow}

\begin{figure}
    \centering
    \includegraphics[page=16, width=\textwidth]{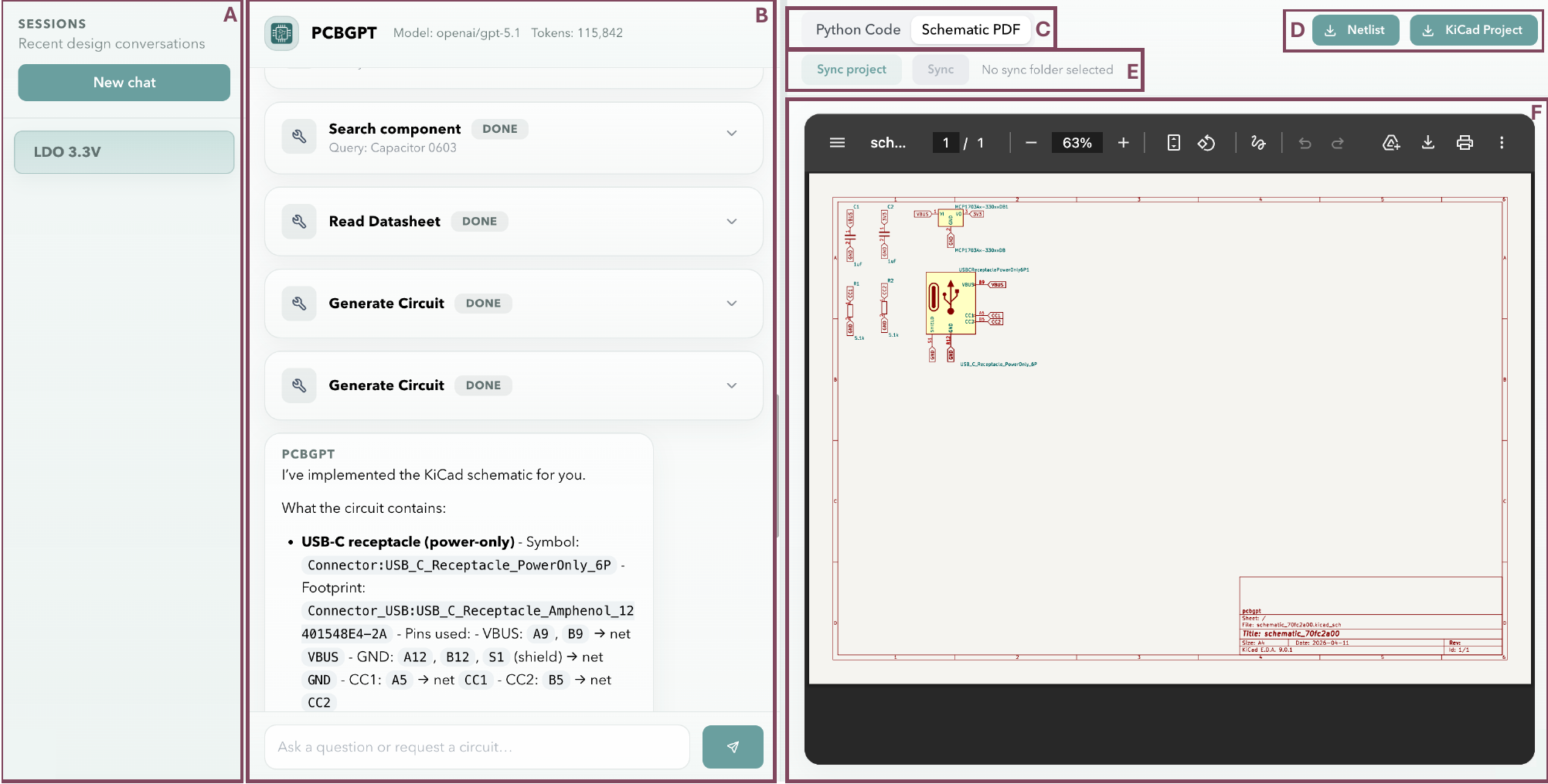}
    \caption{Interactive web workflow showing session management (A), chat and tool traces (B), schematic and DSL views (C), export options (D), local project synchronization (E), and schematic preview (F).}
    \Description{Screenshot-style overview of the web interface. It highlights session management, a chat area with tool traces, schematic and DSL views, export options, local project synchronization, and a rendered schematic preview.}
\end{figure}

%% file: figures/system_overview.tex
\begin{figure*}[t]
\centering
\includegraphics[page=13, width=\textwidth]{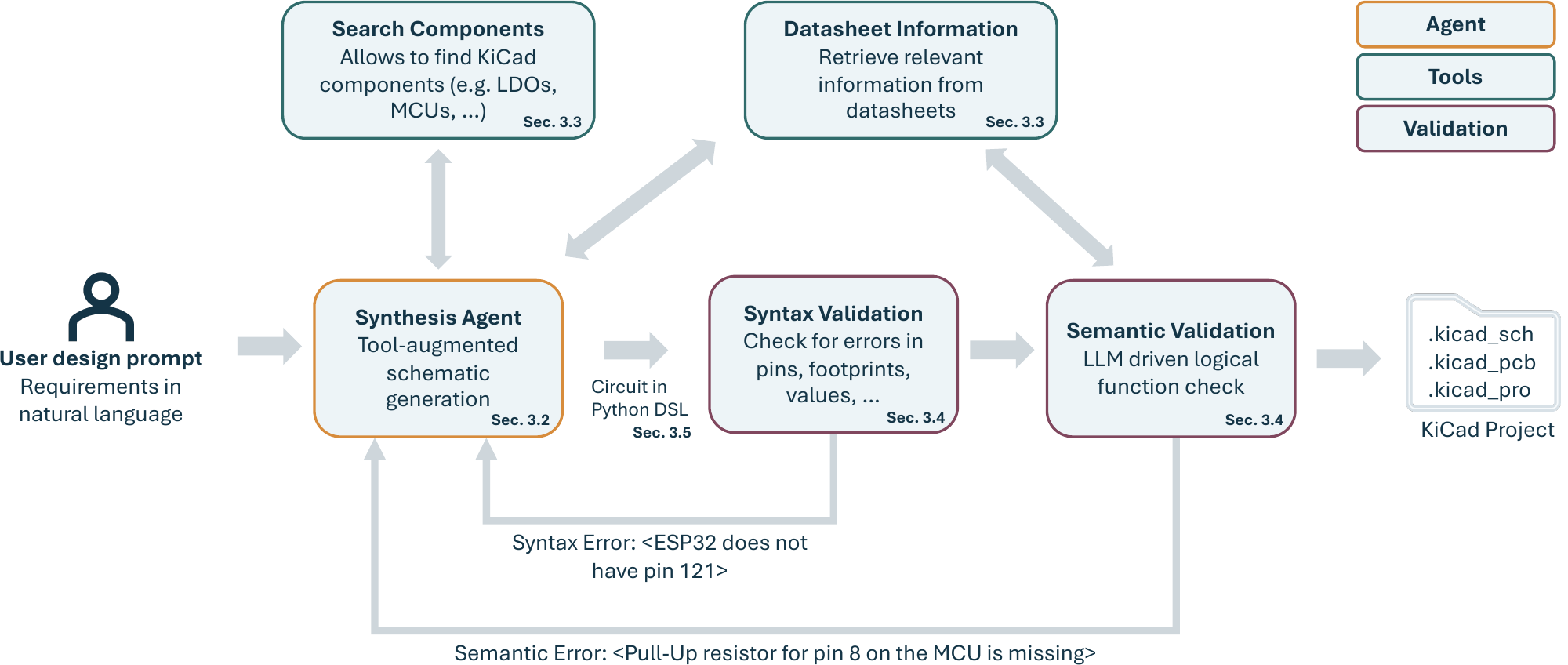}
\caption{System overview of the \systemName{} pipeline. A natural-language design prompt is processed by the synthesis agent, which uses the tool set to ground component and pin-selection decisions. The generated Python circuit-DSL code is passed through the validation pipeline, which performs execution, KiCad export, and validation and returns feedback to the synthesis agent. The synthesis agent then outputs a KiCad project draft together with netlist and PDF artifacts for downstream inspection and refinement.}
\Description{A pipeline overview of pcbGPT. A natural-language design prompt enters a synthesis agent. The synthesis agent uses grounding tools for component and pin selection, produces Python circuit-DSL code, and sends it to a validation pipeline. The validation pipeline performs execution, KiCad export, and validation, then returns feedback to the synthesis agent. The final outputs are a KiCad project draft together with netlist and PDF artifacts for inspection and refinement.}
\label{fig:system-overview}
\end{figure*}

%% file: figures/web_workflow.tex
\begin{figure*}[t]
\centering

\includegraphics[page=14, width=\textwidth]{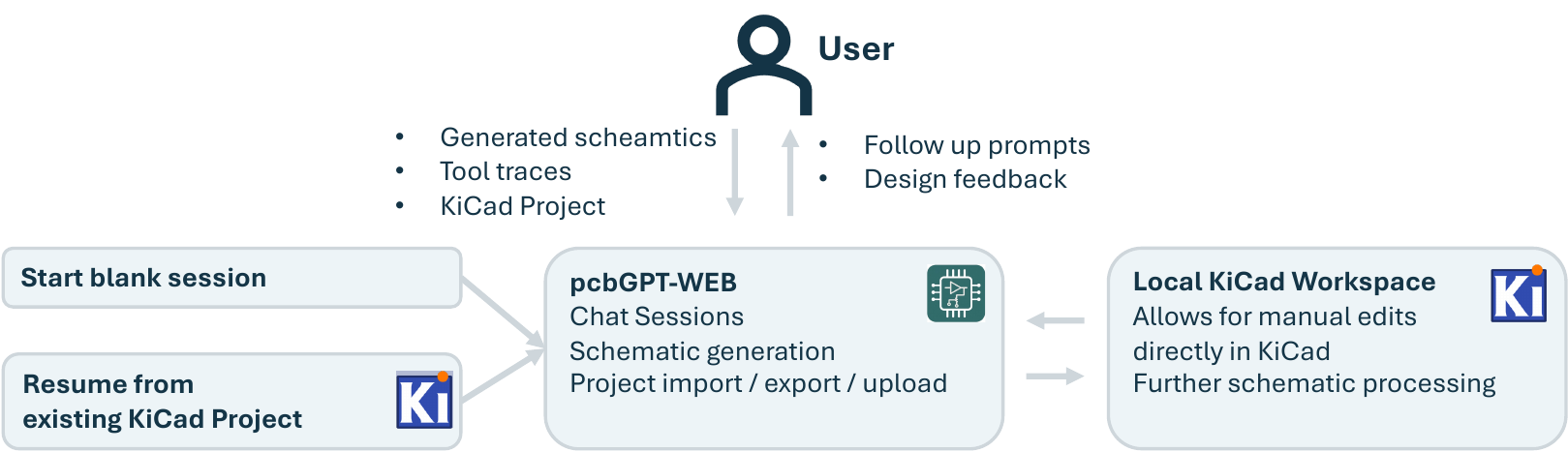}

\caption{Interactive workflow supported by \systemName-Web. A session can begin either from a new natural-language requirements description or from an imported KiCad project. The browser session maintains shared conversational and project context across rounds, while users iteratively inspect generated schematics, review validation feedback, refine prompts, make optional local KiCad edits, and synchronize updated projects back into the system.}
\Description{An overview of the interactive pcbGPT-Web workflow. A session starts either from new natural-language requirements or from an imported KiCad project. The browser session maintains shared conversation history and project context across rounds. Users inspect generated schematics, review validation feedback, refine prompts, optionally edit the project locally in KiCad, and synchronize updated project state back into the web system.}
\label{fig:web-workflow}
\end{figure*}

%% file: chapters/benchmark_design.tex
\subsection{Benchmark Design}
\label{sec:benchmark-design}
To evaluate the performance of \systemName{} systematically, we constructed a set of natural-language design tasks paired with reference implementations and prompt-level interface constraints.
In the current codebase, this task set is represented as a registry of experiment definitions paired with reference circuits, enabling reproducible execution and automatic evaluation.
The task set spans four difficulty tiers: basic, easy, medium, and hard.
These tasks are intentionally oriented toward embedded and wearable electronics, with emphasis on sensing, wireless communication, power management, microcontroller interfacing, and compact battery-powered subsystems.
This focus reflects the target application domain of the synthesis agent and exercises the kinds of multi-component integration challenges that arise in practical human-centered hardware design.

\subsubsection{Benchmark Construction}
\label{sec:benchmark-prompt-construction}

Each benchmark task is defined as a structured experiment specification containing a natural-language description, a difficulty label, a unique identifier, an optional list of required key components, and optional required interface nets.
During experiment execution, these prompts are automatically augmented with benchmark-specific instructions that require the generation agent to expose interface nets with exact names and to keep iterating until code generation succeeds.
This design ensures that evaluation is not limited to free-form textual descriptions, but instead includes explicit task-level constraints that can later be checked automatically.

The active prompt set is organized into four difficulty tiers.
The \emph{basic} tier contains small building blocks such as regulators, filters, and single-transistor stages.
The \emph{easy} tier places one interface or storage block into a small embedded context, for example I\textsuperscript{2}C EEPROM configuration, bidirectional I\textsuperscript{2}C level shifting, SD-card SPI, and SPI flash interfaces.
The \emph{medium} tier introduces multi-device interaction and power-handling tasks such as buck conversion, CAN bus MCU interfacing, USB-C power, battery charging, RTC backup, sensor indication, and I\textsuperscript{2}C multiplexing.
The \emph{hard} tier captures larger subsystem-integration tasks such as audio recording, USB-C-connected CAN sensing, BLE IMU logging, LoRa sensing, and vehicle tracking nodes. Across these tasks, the dominant functional categories are sensing, wireless communication, power conversion and charging, storage, and microcontroller-centric interface design.

This task-set design is motivated by the observation that many PCB schematic generation tasks are not isolated component-selection problems, but integration problems in which multiple subsystems must be combined under interface and power constraints. Accordingly, the prompts do not only ask for isolated circuits; they also require the model to expose specific interfaces such as \texttt{+3V3}, \texttt{GND}, \texttt{SDA}, \texttt{SCL} or \texttt{ADC\_IN}, and in many cases to include concrete KiCad components such as regulators, transceivers, sensors, or MCU modules.
The result is an evaluation set that measures whether the system can synthesize structurally coherent, constraint-satisfying embedded schematics rather than merely plausible isolated fragments.

Concretely, each task in this set contains the following attributes:
\begin{itemize}
    \item \textbf{Id}: A unique integer identifier for the prompt.
    \item \textbf{Prompt}: The natural-language task description given to the system.
    \item \textbf{Short Description}: A concise summary of the task
    \item \textbf{Key Components}: An optional list of required components that must be present in the generated circuit, specified by the KiCad library and symbol name. This helps ensure the generated design can be compared against the reference implementation at the component level.
    \item \textbf{Required Nets}: An optional list of nets which must be present in the generated circuit with exact names. This allows the benchmark to check whether the generated design exposes the required external interfaces.
    \item \textbf{Difficulty}: A categorical label indicating the difficulty tier of the task, which can be used for analysis and stratification of results.
    \item \textbf{Reference Implementation}: A Python function that constructs a reference circuit in the same DSL as the generated outputs, which serves as the target for structural similarity evaluation.
\end{itemize}

\newcommand{\BenchmarkCompactCaption}{Overview of the 20 \systemName{}-Eval benchmark tasks by difficulty tier. In addition to prompt identifiers and short descriptions, the table reports reference-circuit size and connectivity statistics (components, pins, nets, and connected pins) used to characterize benchmark complexity; full natural-language prompt texts are listed in Appendix~\ref{app:benchmark-inventory}.}
\input{tables/benchmark_compact.tex}

An overview of the benchmark problems is provided in \autoref{tab:benchmark-compact}. A full per-task listing with descriptions, required parts, and exposed interface nets is given in Appendix~\ref{app:benchmark-inventory}.

\subsection{Reference Implementations}
The reference implementations for each benchmark are not merely static ground-truth circuits.
Instead, they encode acceptable design variation directly in the circuit specification.
For example, reference circuits may include optional parts when these are identified as optional in the corresponding datasheet or reference design, so that generated solutions are not penalized for omitting components that are not strictly required for correct operation.
In addition, the reference implementations support flexible pin assignments through the \texttt{FlexiblePin} mechanism, which allows multiple valid pin mappings where the underlying component or module permits interchangeable connections.
This makes the benchmark evaluation more robust to superficial variations in pin assignments that do not affect functional correctness, while still enforcing the presence of required components and interface nets.
For example, the use of GPIO-pins on a microcontroller may be flexible, but the presence of the microcontroller itself and the exposure of required I/O nets are not.

These reference implementations are datasheet-grounded exemplar solutions, not exhaustive descriptions of the full design space, especially for larger tasks where multiple electrically valid schematic realizations may exist.
Although the benchmark allows some flexibility through optional parts, optional values, and flexible pin assignments, it does not cover all acceptable variants, so non-matches should be interpreted conservatively rather than as proof that a generated circuit is invalid in practice.

\input{chapters/reference_first_global_comparator.tex}

%% file: tables/benchmark_compact.tex
\begin{table*}[t]
\centering
\small
\caption{\BenchmarkCompactCaption}
\label{tab:benchmark-compact}
\setlength{\tabcolsep}{4pt}
\begin{tabularx}{\textwidth}{@{}crlrrrrX@{}}
\toprule
Tier & Id & Identifier & Comp. & Pins & Nets & Conn. pins & Circuit description \\
\midrule
\rowcolor{green!12}
 & 1 & ADCFE & 2 & 4 & 3 & 4 & ADC front-end conditioning \\
\rowcolor{green!12}
 & 2 & LDO33 & 5 & 11 & 3 & 11 & 5V-to-3.3V LDO regulator \\
\rowcolor{green!12}
 & 3 & NPNLS & 3 & 7 & 5 & 7 & NPN open-collector level shifter \\
\rowcolor{green!12}
\multirow{-4}{*}{\rotatebox[origin=c]{90}{\textbf{Basic}}} & 4 & RC1K & 2 & 4 & 3 & 4 & 1 kHz RC low-pass \\
\rowcolor{yellow!20}
 & 5 & EECFG & 5 & 44 & 4 & 19 & I2C EEPROM configuration \\
\rowcolor{yellow!20}
 & 6 & I2CLVL & 6 & 14 & 6 & 14 & Bidirectional I2C level shifter \\
\rowcolor{yellow!20}
 & 7 & SDSPI & 6 & 48 & 6 & 22 & SD card SPI interface \\
\rowcolor{yellow!20}
\multirow{-4}{*}{\rotatebox[origin=c]{90}{\textbf{Easy}}} & 8 & SPIFLS & 5 & 44 & 8 & 21 & SPI flash memory interface \\
\rowcolor{orange!18}
 & 9 & BUCK5 & 12 & 31 & 7 & 29 & 12V-to-5V buck converter \\
\rowcolor{orange!18}
 & 10 & CANBUS & 8 & 84 & 9 & 32 & CAN bus MCU interface \\
\rowcolor{orange!18}
 & 11 & CCSLED & 12 & 61 & 12 & 37 & CCS811 air-quality indicator \\
\rowcolor{orange!18}
 & 12 & I2CMUX & 13 & 81 & 9 & 44 & I2C mux dual same-address sensors \\
\rowcolor{orange!18}
 & 13 & LICHRG & 10 & 32 & 8 & 31 & USB-C Li-Ion charger \\
\rowcolor{orange!18}
 & 14 & RTCBKP & 6 & 54 & 6 & 28 & RTC backup-domain interface \\
\rowcolor{orange!18}
\multirow{-7}{*}{\rotatebox[origin=c]{90}{\textbf{Medium}}} & 15 & USBC3V & 8 & 22 & 6 & 22 & USB-C to 3.3V supply \\
\rowcolor{red!15}
 & 16 & AUDREC & 13 & 70 & 11 & 50 & Portable audio recorder \\
\rowcolor{red!15}
 & 17 & ICAN & 23 & 115 & 15 & 82 & USB CAN sensor node \\
\rowcolor{red!15}
 & 18 & IMULOG & 34 & 132 & 23 & 113 & BLE IMU data logger \\
\rowcolor{red!15}
 & 19 & LORASEN & 32 & 128 & 21 & 108 & LoRa sensor node \\
\rowcolor{red!15}
\multirow{-5}{*}{\rotatebox[origin=c]{90}{\textbf{Hard}}} & 20 & VTRACK & 28 & 136 & 19 & 92 & Vehicle tracking node \\
\bottomrule
\end{tabularx}
\end{table*}

%% file: chapters/reference_first_global_comparator.tex
\subsection{Circuit Comparison}
\label{sec:reference-first-global-comparator}
To compare the generated circuits against the reference solutions, we employ a deterministic comparator which works by matching components and nets between the generated and reference circuits.
Based on that matching, the comparator computes a similarity score that reflects how closely the generated circuit matches the reference in terms of components, attributes, and connectivity.
The algorithm works in three main stages: component matching, net matching, and score computation.
The algorithm is summarized in \autoref{alg:reference-first-global}.

\subsubsection{Component Matching}
Let the reference circuit be $A=(C_R,N_R)$ and the generated circuit be $B=(C_G,N_G)$, where $C$ denotes the set of components and $N$ denotes the set of nets.
Using the flags in the reference circuit, we partition the reference components into required and optional sets:
\[
C_R = C_R^{\mathrm{req}} \cup C_R^{\mathrm{opt}},
\qquad
C_R^{\mathrm{req}} \cap C_R^{\mathrm{opt}} = \varnothing.
\]
Matching is then performed in two passes:
\begin{enumerate}
    \item first match all required reference components $C_R^{\mathrm{req}}$ to generated components,
    \item then match optional reference components $C_R^{\mathrm{opt}}$ only against the remaining unmatched generated components.
\end{enumerate}
Only components with identical KiCad library and symbol name are considered compatible candidates.
Within each compatible candidate set, the comparator computes a connection-sensitive weight and then solves a one-to-one assignment using the Hungarian algorithm to resolve ambiguities among repeated plausible matches such as pull-ups or decoupling capacitors.
This ensures that the selected assignment is globally consistent rather than determined by local or iteration-order effects.

For a reference component $c_r$ and generated component $c_g$, connection similarity is computed from pin-level local neighborhood signatures: each pin is represented by the multiset of neighboring component-library, symbol-name, and pin tuples connected to the same net, excluding the component itself.
The reference and generated pin signatures are then aligned one-to-one, and candidates with greater structural overlap are preferred.
Value agreement is used only as a final tie-breaker after normalizing component values, including tolerance-based comparison for numeric passive values when applicable.

When several components share the same symbolic type and nominal value, such as repeated pull-ups or decoupling capacitors, multiple plausible matches may exist.
The comparator therefore resolves these ambiguities by choosing the global one-to-one assignment that maximizes structural agreement with the reference.

\paragraph{Optional components and optional nets.}
Let $M$ denote the reference-to-generated component mapping produced by the two-pass assignment.
The effective reference side used for scoring is then
\[
\widetilde{C}_R
=
C_R^{\mathrm{req}}
\cup
\{\, c \in C_R^{\mathrm{opt}} \mid c \in \mathrm{dom}(M) \,\},
\]
that is, all required reference components plus only those optional reference components that were actually matched.
Unmatched optional reference components are ignored completely: they do not appear as missing components and they do not enlarge the denominator of the component score.

The same idea is applied to optional reference nets.
If a net is marked optional in the reference and is absent from the generated circuit, the comparator does not penalize that absence.
If it is present, however, it participates in connectivity comparison like any other scored net.

\paragraph{Endpoint normalization.}
Connectivity is evaluated on sets of component-pin endpoints.
For each scored net $n$, let $E(n)$ denote the set of connected component-pin endpoints after replacing each generated component reference by its matched reference component, when such a match exists.
Two normalizations are used:
\begin{enumerate}
    \item symmetric two-pin passives such as resistors, inductors, and non-polarized capacitors are canonicalized to a synthetic pin label \texttt{\_\_sym\_\_}, so pin-order swaps are not penalized;
    \item flexible reference pins are expanded into endpoint variants, so one reference endpoint may match any of a declared set of alternative MCU or module pins.
\end{enumerate}

\subsubsection{Scoring}
After component matching, the comparator computes three subscores: a component score, an attribute score, and a connectivity score.
These are then combined into the final similarity score.

\paragraph{Component score.}
Let $M$ denote the reference-to-generated component mapping, let $\widetilde{C}_R$ be the scored reference component set after optional-component filtering, and let $C_G$ be the generated component set.
The component score is the Dice coefficient
\[
S_{\mathrm{comp}} = \frac{2|M|}{|\widetilde{C}_R| + |C_G|}.
\]
This rewards overlap in the component sets while penalizing both missing required reference components and extra generated components.

\paragraph{Attribute score.}
For each matched component pair, the comparator records attribute differences after component matching.
Because matching only permits pairs with identical library and symbol name, the only remaining attribute difference in practice is the component value.
The attribute score is therefore determined by value agreement across matched pairs.

Let $D$ be the number of matched pairs whose values do not agree, and let $|M|$ be the number of matched component pairs.
Then
\[
S_{\mathrm{attr}} =
\begin{cases}
1, & \text{if } |M| = 0,\\[4pt]
1 - \frac{D}{|M|}, & \text{otherwise.}
\end{cases}
\]
For passive components, the implementation normalizes common numeric forms and supports tolerance-based comparison, so equivalent or near-equivalent values can still count as matches.

\paragraph{Connectivity score.}
Connectivity is evaluated on endpoint sets after applying the component mapping.
For a matched pair of nets $(n_i,m_j)$, let
\[
m_{ij} = |E(n_i) \cap E(m_j)|,
\qquad
u_{ij} = |E(n_i)| + |E(m_j)| - m_{ij},
\]
where $E(n)$ denotes the endpoint set of net $n$ after endpoint normalization.
Let $\mathcal{M}_N$ denote the net assignment chosen by the comparator.
The connectivity score is
\[
S_{\mathrm{conn}} =
\frac{\sum_{(i,j)\in\mathcal{M}_N} m_{ij}}
{\sum_{(i,j)\in\mathcal{M}_N} u_{ij}
\;+\;
\sum_{n \in N_R^{\mathrm{unmatched,req}}} |E(n)|
\;+\;
\sum_{m \in N_G^{\mathrm{unmatched}}} |E(m)| }.
\]
Thus, matched nets contribute according to endpoint overlap, unmatched required reference nets contribute their full endpoint count to the denominator, and unmatched generated nets are penalized in the same way, while unmatched optional reference nets are ignored.

\paragraph{Final similarity.}
The final similarity score is a weighted combination of the three subscores:
\[
S = \alpha S_{\mathrm{comp}} + \beta S_{\mathrm{attr}} + \gamma S_{\mathrm{conn}},
\]
with default weights
\[
(\alpha,\beta,\gamma) = (0.4, 0.2, 0.4).
\]

\input{figures/reference_first_global_pseudocode.tex}

%% file: figures/reference_first_global_pseudocode.tex
\begin{algorithm}[t]
\caption{Reference-first global circuit comparison}
\label{alg:reference-first-global}
\KwIn{reference circuit $A$, generated circuit $B$}
\KwOut{similarity score $S$}

$M \gets \varnothing$ \tcp*{reference $\rightarrow$ generated component mapping}

$C_R^{\mathrm{req}}, C_R^{\mathrm{opt}} \gets$ split reference components into required and optional\;
$M \gets \mathrm{HungarianMatch}(C_R^{\mathrm{req}}, C_G)$ using compatible pairs and structural weights\;
$U \gets \mathrm{image}(M)$\;
$M \gets M \cup \mathrm{HungarianMatch}(C_R^{\mathrm{opt}}, C_G \setminus U)$ using the same weights\;

$\widetilde{C}_R \gets C_R^{\mathrm{req}} \cup \{c \in C_R^{\mathrm{opt}} \mid c \in \mathrm{dom}(M)\}$\;
$S_{\mathrm{comp}} \gets \dfrac{2|M|}{|\widetilde{C}_R| + |C_G|}$\;

$D \gets$ number of matched pairs with non-matching normalized values\;
$S_{\mathrm{attr}} \gets
\begin{cases}
1, & \text{if } |M| = 0,\\
1 - \dfrac{D}{|M|}, & \text{otherwise}
\end{cases}$\;

Construct normalized endpoint sets $E_R(n)$ for $n \in N_R$ and $E_G(m)$ for $m \in N_G$ using $M$\;
$\mathcal{M}_N \gets \mathrm{HungarianMatch}(N_R, N_G)$ with weights from $m_{ij} = |E_R(n_i) \cap E_G(m_j)|$ and $u_{ij} = |E_R(n_i)| + |E_G(m_j)| - m_{ij}$\;
Ignore unmatched optional reference nets\;
$S_{\mathrm{conn}} \gets
\dfrac{
\sum_{(i,j)\in \mathcal{M}_N} m_{ij}
}{
\sum_{(i,j)\in \mathcal{M}_N} u_{ij}
\;+\;
\sum_{n \in N_R^{\mathrm{unmatched,req}}} |E(n)|
\;+\;
\sum_{m \in N_G^{\mathrm{unmatched}}} |E(m)|
}$\;

$S \gets 0.4 S_{\mathrm{comp}} + 0.2 S_{\mathrm{attr}} + 0.4 S_{\mathrm{conn}}$\;
\Return{$S$}\;
\end{algorithm}

%% file: chapters/experiments.tex
\section{Experiments}
\systemName{} supports both fully automatic schematic generation and interactive human-in-the-loop use through the interactive web workflow. In this section, we evaluate the automatic mode to answer three questions: whether the system can generate correct editable schematics from natural-language prompts, how performance changes with task complexity, and whether the automatic comparator aligns with expert judgment.

\subsection{Experimental Setup}
To answer these questions, we conduct an automatic evaluation on 20 representative embedded, IoT, and wearable design tasks and compare generated schematics against benchmark reference designs.

To evaluate the capability of \systemName{} to generate correct schematics from natural-language prompts, we employ large language models from both commercial and open-source providers.
Specifically, we evaluate GPT-5.1 (gpt-5.1-2025-11-13) \cite{GPT51:online} and GPT-5.3-Codex (gpt-5.3-codex) \cite{GPT53Codex:online} from OpenAI, as well as Qwen3.5-35B-A3B \cite{qwen3.5} and Qwen3.5-397B-A17B \cite{qwen3.5} from Alibaba.
For each task-model pair, we execute 5 independent runs.
In comparison to the interactive web version, the runner augments every prompt with two fixed instructions: the agent must expose all required interface nets with exact names, and it must keep iterating until a circuit is produced.

For one experiment run, we use the same model for the synthesis, generation tools, and post-generation validation stages.
For both generation and validation we set the temperature of the LLMs to 0.2, which is a common setting for code generation tasks.
We evaluate only vision-capable models because the current grounding and validation pipeline relies on visual datasheet understanding during component retrieval and review.

\subsection{Automatic Evaluation Metric}
We report two automatic evaluation metrics derived from the comparator described in \autoref{sec:reference-first-global-comparator}: a continuous similarity score for each generated circuit and a pass@k score derived from exact matches.
For each generated candidate, we first compute the comparator similarity, which measures how closely the generated circuit matches the reference in terms of components, attributes, and connectivity.
A run is counted as successful for pass@k if and only if its comparator similarity is 1.0, that is, if the generated circuit exactly matches the reference under the comparator's normalization rules for optional components, flexible pins, and equivalent connectivity patterns.
Given the number of successful runs for a prompt-model pair, we then compute pass@k using the standard estimator over the sampled candidates.
This success criterion is intentionally conservative. The benchmark references are datasheet-grounded exemplar solutions rather than an exhaustive encoding of all valid circuits, so an exact comparator match should be read as strong evidence of correctness, while non-matches can still include practically useful or even acceptable alternative realizations.

\subsection{Results}
Tables~\ref{tab:automatic-results-pass-at} and \ref{tab:automatic-results-average-similarity} summarize the pass@1, pass@3, pass@5, and average-similarity results for each model and task combination in our evaluation set.

Several conclusions are relevant for practical utility.
First, the evaluation shows that the system can already solve a substantial portion of early-stage schematic tasks automatically, especially at the basic, easy, and medium difficulty levels.
\AutomaticResultsBestModelName{} is the strongest model overall, with an overall pass@1 of \red{\AutomaticResultsBestOverallPassAtOne{}} and pass@5 of \red{\AutomaticResultsBestOverallPassAtFive{}}.
It solves all basic and easy tasks at pass@1 = \red{\AutomaticResultsBestBasicPassAtOne{}}, remains strong on medium tasks with pass@1 = \red{\AutomaticResultsBestMediumPassAtOne{}}, and degrades mainly on the hard tasks where pass@1 drops to \red{\AutomaticResultsBestHardPassAtOne{}}.
The second-best model, \AutomaticResultsSecondModelName{}, reaches an overall pass@1 of \red{\AutomaticResultsSecondOverallPassAtOne{}} and pass@5 of \red{\AutomaticResultsSecondOverallPassAtFive{}}, with pass@1 = \red{\AutomaticResultsSecondBasicPassAtOne{}} on both the basic and easy tiers, \red{\AutomaticResultsSecondMediumPassAtOne{}} on medium tasks, and \red{\AutomaticResultsSecondHardPassAtOne{}} on hard tasks.
This pattern suggests that once the task requires coordinating several subsystems, the dominant failure mode is no longer local symbol or wiring mistakes, but maintaining global consistency across power, interfaces, and support circuitry.
Overall, this shows that automatic schematic generation is already reliable for simpler tasks and partially reliable for larger subsystem designs.

Second, repeated sampling is practically meaningful.
For the stronger models, pass@5 is consistently higher than pass@1 on the more difficult tasks, reaching \red{\AutomaticResultsBestMediumPassAtFive{}} on medium tasks and \red{\AutomaticResultsBestHardPassAtFive{}} on hard tasks even when single-shot performance is clearly lower.
This means the system is better interpreted as a candidate generator that benefits from reranking or human review than as a deterministic one-shot schematic synthesizer.
In other words, sampling recovers many correct designs that are within reach of the model distribution but not reliably produced on the first attempt.
This indicates that the automatic mode is especially useful for producing reviewable candidate schematics rather than only one final answer.

Third, Table~\ref{tab:automatic-results-average-similarity} shows that pass@k alone would understate how close many failed generations are to the reference.
This is particularly visible on the hard tasks: for \AutomaticResultsBestModelName{} the hard-task mean similarity rounds to \AutomaticResultsBestHardAverageSimilarity{} while hard-task pass@1 is only \red{\AutomaticResultsBestHardPassAtOne{}}, and even models with poor hard-task pass rates often remain very high in average similarity.
The implication is that many failures are near misses rather than complete breakdowns.
Typical examples are circuits that capture the correct subsystem decomposition and most connectivity, but miss one required net exposure, omit a supporting passive, or realize a constraint in a way that is structurally close yet not equivalent under the exact-match criterion.
This gap becomes more important as task complexity increases, because larger embedded design problems admit a broader design space and often allow multiple correct solutions.
Our benchmark references are intentionally datasheet-grounded and allow some flexibility, but they still represent specific reviewed realizations rather than an exhaustive set of all acceptable circuits for a prompt.
This distinction matters because, from an engineering workflow perspective, these near misses may still be useful starting points for expert refinement even though they do not count as benchmark successes.
This suggests that exact-match pass@k is a strict measure of automatic success, but not a complete measure of practical usefulness.

Finally, the comparison to human expert grading supports the validity of the automatic comparator while also clarifying its bias.
Across \HumanComparatorEvaluatedRuns{} evaluated runs, human and automatic labels agree in \HumanComparatorAgreementPct{}\% of cases, with Cohen's $\kappa = \HumanComparatorKappa{}$, which indicates strong agreement beyond chance.
More importantly, the disagreement pattern is asymmetric: the comparator produces no false positives and \HumanComparatorFalseNegative{} false negatives relative to the human labels.
Thus, when the comparator declares a design correct, it is highly trustworthy; when it rejects a design, the output is sometimes still considered working by the expert but falls short of the comparator's exact-match requirement.
This makes the comparator suitable for conservative benchmark reporting: it avoids overstating performance, but it can underestimate the practical usefulness of partially correct or acceptably alternative designs.
This supports using the comparator for automatic evaluation while still interpreting failures conservatively.

Taken together, the results indicate that \systemName{} is already effective for drafting and exploring many small to medium embedded schematics, especially when multiple candidates can be sampled.
However, the hard-task results and the gap between similarity and pass@k also show that fully autonomous generation of complex multi-subsystem circuits remains unreliable.
The current system is therefore best positioned as a design acceleration tool for hardware-literate users and experts rather than as a replacement for expert schematic review.

\input{tables/automatic_results_pass_at.tex}

\input{tables/automatic_results_average_similarity.tex}

%% file: tables/automatic_results_pass_at.tex
\begin{table*}[t]
\centering
\small
\caption{Per-problem pass@k scores across evaluated models using deterministic global structural similarity.}
\label{tab:automatic-results-pass-at}
\setlength{\tabcolsep}{4pt}
\resizebox{\linewidth}{!}{%
\begin{tabular}{l|rrr|rrr|rrr|rrr}
\toprule
\textbf{Identifier} & \multicolumn{3}{|c}{\textbf{gpt-5.3-codex}} & \multicolumn{3}{|c}{\textbf{qwen3.5-397b-a17b}} & \multicolumn{3}{|c}{\textbf{gpt-5.1}} & \multicolumn{3}{|c}{\textbf{qwen3.5-35b-a3b}} \\
 & \textbf{Pass@1} & \textbf{Pass@3} & \textbf{Pass@5} & \textbf{Pass@1} & \textbf{Pass@3} & \textbf{Pass@5} & \textbf{Pass@1} & \textbf{Pass@3} & \textbf{Pass@5} & \textbf{Pass@1} & \textbf{Pass@3} & \textbf{Pass@5} \\
\midrule
\rowcolor{green!12}
ADCFE & 1.00 & 1.00 & 1.00 & 1.00 & 1.00 & 1.00 & 0.60 & 1.00 & 1.00 & 1.00 & 1.00 & 1.00 \\
\rowcolor{green!12}
LDO33 & 1.00 & 1.00 & 1.00 & 1.00 & 1.00 & 1.00 & 0.80 & 1.00 & 1.00 & 1.00 & 1.00 & 1.00 \\
\rowcolor{green!12}
NPNLS & 1.00 & 1.00 & 1.00 & 1.00 & 1.00 & 1.00 & 1.00 & 1.00 & 1.00 & 1.00 & 1.00 & 1.00 \\
\rowcolor{green!12}
RC1K & 1.00 & 1.00 & 1.00 & 1.00 & 1.00 & 1.00 & 1.00 & 1.00 & 1.00 & 1.00 & 1.00 & 1.00 \\
\rowcolor{green!12}
\textbf{Basic Avg.} & \textbf{1.00} & \textbf{1.00} & \textbf{1.00} & \textbf{1.00} & \textbf{1.00} & \textbf{1.00} & \textbf{0.85} & \textbf{1.00} & \textbf{1.00} & \textbf{1.00} & \textbf{1.00} & \textbf{1.00} \\
\midrule
\rowcolor{yellow!20}
EECFG & 1.00 & 1.00 & 1.00 & 1.00 & 1.00 & 1.00 & 1.00 & 1.00 & 1.00 & 1.00 & 1.00 & 1.00 \\
\rowcolor{yellow!20}
I2CLVL & 1.00 & 1.00 & 1.00 & 1.00 & 1.00 & 1.00 & 1.00 & 1.00 & 1.00 & 0.60 & 1.00 & 1.00 \\
\rowcolor{yellow!20}
SDSPI & 1.00 & 1.00 & 1.00 & 1.00 & 1.00 & 1.00 & 1.00 & 1.00 & 1.00 & 0.20 & 0.60 & 1.00 \\
\rowcolor{yellow!20}
SPIFLS & 1.00 & 1.00 & 1.00 & 1.00 & 1.00 & 1.00 & 0.80 & 1.00 & 1.00 & 0.80 & 1.00 & 1.00 \\
\rowcolor{yellow!20}
\textbf{Easy Avg.} & \textbf{1.00} & \textbf{1.00} & \textbf{1.00} & \textbf{1.00} & \textbf{1.00} & \textbf{1.00} & \textbf{0.95} & \textbf{1.00} & \textbf{1.00} & \textbf{0.65} & \textbf{0.90} & \textbf{1.00} \\
\midrule
\rowcolor{orange!18}
BUCK5 & 0.80 & 1.00 & 1.00 & 0.40 & 0.90 & 1.00 & 0.20 & 0.60 & 1.00 & 0.00 & 0.00 & 0.00 \\
\rowcolor{orange!18}
CANBUS & 0.60 & 1.00 & 1.00 & 0.80 & 1.00 & 1.00 & 0.60 & 1.00 & 1.00 & 0.60 & 1.00 & 1.00 \\
\rowcolor{orange!18}
CCSLED & 1.00 & 1.00 & 1.00 & 0.80 & 1.00 & 1.00 & 0.80 & 1.00 & 1.00 & 0.60 & 1.00 & 1.00 \\
\rowcolor{orange!18}
I2CMUX & 1.00 & 1.00 & 1.00 & 0.80 & 1.00 & 1.00 & 0.80 & 1.00 & 1.00 & 0.80 & 1.00 & 1.00 \\
\rowcolor{orange!18}
LICHRG & 1.00 & 1.00 & 1.00 & 0.60 & 1.00 & 1.00 & 0.40 & 0.90 & 1.00 & 0.60 & 1.00 & 1.00 \\
\rowcolor{orange!18}
RTCBKP & 1.00 & 1.00 & 1.00 & 1.00 & 1.00 & 1.00 & 0.80 & 1.00 & 1.00 & 0.80 & 1.00 & 1.00 \\
\rowcolor{orange!18}
USBC3V & 1.00 & 1.00 & 1.00 & 0.60 & 1.00 & 1.00 & 0.60 & 1.00 & 1.00 & 0.60 & 1.00 & 1.00 \\
\rowcolor{orange!18}
\textbf{Medium Avg.} & \textbf{0.91} & \textbf{1.00} & \textbf{1.00} & \textbf{0.71} & \textbf{0.99} & \textbf{1.00} & \textbf{0.60} & \textbf{0.93} & \textbf{1.00} & \textbf{0.57} & \textbf{0.86} & \textbf{0.86} \\
\midrule
\rowcolor{red!15}
AUDREC & 1.00 & 1.00 & 1.00 & 1.00 & 1.00 & 1.00 & 0.60 & 1.00 & 1.00 & 0.00 & 0.00 & 0.00 \\
\rowcolor{red!15}
ICAN & 0.20 & 0.60 & 1.00 & 0.00 & 0.00 & 0.00 & 0.00 & 0.00 & 0.00 & 0.00 & 0.00 & 0.00 \\
\rowcolor{red!15}
IMULOG & 0.80 & 1.00 & 1.00 & 0.00 & 0.00 & 0.00 & 0.00 & 0.00 & 0.00 & 0.00 & 0.00 & 0.00 \\
\rowcolor{red!15}
LORASEN & 1.00 & 1.00 & 1.00 & 0.20 & 0.60 & 1.00 & 0.00 & 0.00 & 0.00 & 0.00 & 0.00 & 0.00 \\
\rowcolor{red!15}
VTRACK & 0.60 & 1.00 & 1.00 & 0.20 & 0.60 & 1.00 & 0.00 & 0.00 & 0.00 & 0.00 & 0.00 & 0.00 \\
\rowcolor{red!15}
\textbf{Hard Avg.} & \textbf{0.72} & \textbf{0.92} & \textbf{1.00} & \textbf{0.28} & \textbf{0.44} & \textbf{0.60} & \textbf{0.12} & \textbf{0.20} & \textbf{0.20} & \textbf{0.00} & \textbf{0.00} & \textbf{0.00} \\
\midrule
\textbf{Overall Avg.} & \textbf{0.90} & \textbf{0.98} & \textbf{1.00} & \textbf{0.72} & \textbf{0.86} & \textbf{0.90} & \textbf{0.60} & \textbf{0.78} & \textbf{0.80} & \textbf{0.53} & \textbf{0.68} & \textbf{0.70} \\
\bottomrule
\end{tabular}
}
\end{table*}

%% file: tables/automatic_results_average_similarity.tex
\begin{table*}[t]
\centering
\small
\caption{Per-problem average deterministic global structural similarity across evaluated models.}
\label{tab:automatic-results-average-similarity}
\setlength{\tabcolsep}{4pt}
\resizebox{\linewidth}{!}{%
\begin{tabular}{l|rrrr|rrrr|rrrr|rrrr}
\toprule
\textbf{Identifier} & \multicolumn{4}{|c}{\textbf{gpt-5.3-codex}} & \multicolumn{4}{|c}{\textbf{qwen3.5-397b-a17b}} & \multicolumn{4}{|c}{\textbf{gpt-5.1}} & \multicolumn{4}{|c}{\textbf{qwen3.5-35b-a3b}} \\
 & \textbf{Mean} & \textbf{Std} & \textbf{Min} & \textbf{Max} & \textbf{Mean} & \textbf{Std} & \textbf{Min} & \textbf{Max} & \textbf{Mean} & \textbf{Std} & \textbf{Min} & \textbf{Max} & \textbf{Mean} & \textbf{Std} & \textbf{Min} & \textbf{Max} \\
\midrule
\rowcolor{green!12}
ADCFE & 1.00 & 0.00 & 1.00 & 1.00 & 1.00 & 0.00 & 1.00 & 1.00 & 0.97 & 0.03 & 0.93 & 1.00 & 1.00 & 0.00 & 1.00 & 1.00 \\
\rowcolor{green!12}
LDO33 & 1.00 & 0.00 & 1.00 & 1.00 & 1.00 & 0.00 & 1.00 & 1.00 & 1.00 & 0.01 & 0.98 & 1.00 & 1.00 & 0.00 & 1.00 & 1.00 \\
\rowcolor{green!12}
NPNLS & 1.00 & 0.00 & 1.00 & 1.00 & 1.00 & 0.00 & 1.00 & 1.00 & 1.00 & 0.00 & 1.00 & 1.00 & 1.00 & 0.00 & 1.00 & 1.00 \\
\rowcolor{green!12}
RC1K & 1.00 & 0.00 & 1.00 & 1.00 & 1.00 & 0.00 & 1.00 & 1.00 & 1.00 & 0.00 & 1.00 & 1.00 & 1.00 & 0.00 & 1.00 & 1.00 \\
\rowcolor{green!12}
\textbf{Basic Avg.} & \textbf{1.00} & \textbf{0.00} & \textbf{1.00} & \textbf{1.00} & \textbf{1.00} & \textbf{0.00} & \textbf{1.00} & \textbf{1.00} & \textbf{0.99} & \textbf{0.01} & \textbf{0.97} & \textbf{1.00} & \textbf{1.00} & \textbf{0.00} & \textbf{1.00} & \textbf{1.00} \\
\midrule
\rowcolor{yellow!20}
EECFG & 1.00 & 0.00 & 1.00 & 1.00 & 1.00 & 0.00 & 1.00 & 1.00 & 1.00 & 0.00 & 1.00 & 1.00 & 1.00 & 0.00 & 1.00 & 1.00 \\
\rowcolor{yellow!20}
I2CLVL & 1.00 & 0.00 & 1.00 & 1.00 & 1.00 & 0.00 & 1.00 & 1.00 & 1.00 & 0.00 & 1.00 & 1.00 & 0.93 & 0.09 & 0.76 & 1.00 \\
\rowcolor{yellow!20}
SDSPI & 1.00 & 0.00 & 1.00 & 1.00 & 1.00 & 0.00 & 1.00 & 1.00 & 1.00 & 0.00 & 1.00 & 1.00 & 0.94 & 0.05 & 0.85 & 1.00 \\
\rowcolor{yellow!20}
SPIFLS & 1.00 & 0.00 & 1.00 & 1.00 & 1.00 & 0.00 & 1.00 & 1.00 & 0.99 & 0.02 & 0.94 & 1.00 & 0.99 & 0.02 & 0.94 & 1.00 \\
\rowcolor{yellow!20}
\textbf{Easy Avg.} & \textbf{1.00} & \textbf{0.00} & \textbf{1.00} & \textbf{1.00} & \textbf{1.00} & \textbf{0.00} & \textbf{1.00} & \textbf{1.00} & \textbf{1.00} & \textbf{0.00} & \textbf{0.99} & \textbf{1.00} & \textbf{0.96} & \textbf{0.03} & \textbf{0.93} & \textbf{1.00} \\
\midrule
\rowcolor{orange!18}
BUCK5 & 0.97 & 0.07 & 0.83 & 1.00 & 0.96 & 0.03 & 0.94 & 1.00 & 0.96 & 0.04 & 0.92 & 1.00 & 0.88 & 0.06 & 0.76 & 0.93 \\
\rowcolor{orange!18}
CANBUS & 0.99 & 0.01 & 0.97 & 1.00 & 0.99 & 0.01 & 0.98 & 1.00 & 0.98 & 0.02 & 0.94 & 1.00 & 0.97 & 0.03 & 0.92 & 1.00 \\
\rowcolor{orange!18}
CCSLED & 1.00 & 0.00 & 1.00 & 1.00 & 1.00 & 0.00 & 0.99 & 1.00 & 1.00 & 0.00 & 0.99 & 1.00 & 0.97 & 0.04 & 0.92 & 1.00 \\
\rowcolor{orange!18}
I2CMUX & 1.00 & 0.00 & 1.00 & 1.00 & 0.99 & 0.02 & 0.96 & 1.00 & 0.99 & 0.02 & 0.96 & 1.00 & 0.99 & 0.01 & 0.97 & 1.00 \\
\rowcolor{orange!18}
LICHRG & 1.00 & 0.00 & 1.00 & 1.00 & 0.92 & 0.10 & 0.79 & 1.00 & 0.97 & 0.02 & 0.95 & 1.00 & 0.95 & 0.07 & 0.81 & 1.00 \\
\rowcolor{orange!18}
RTCBKP & 1.00 & 0.00 & 1.00 & 1.00 & 1.00 & 0.00 & 1.00 & 1.00 & 0.99 & 0.01 & 0.97 & 1.00 & 0.97 & 0.05 & 0.87 & 1.00 \\
\rowcolor{orange!18}
USBC3V & 1.00 & 0.00 & 1.00 & 1.00 & 0.95 & 0.07 & 0.87 & 1.00 & 0.96 & 0.08 & 0.80 & 1.00 & 0.95 & 0.07 & 0.87 & 1.00 \\
\rowcolor{orange!18}
\textbf{Medium Avg.} & \textbf{0.99} & \textbf{0.01} & \textbf{0.97} & \textbf{1.00} & \textbf{0.97} & \textbf{0.03} & \textbf{0.92} & \textbf{1.00} & \textbf{0.98} & \textbf{0.02} & \textbf{0.96} & \textbf{1.00} & \textbf{0.96} & \textbf{0.03} & \textbf{0.88} & \textbf{0.99} \\
\midrule
\rowcolor{red!15}
AUDREC & 1.00 & 0.00 & 1.00 & 1.00 & 1.00 & 0.00 & 1.00 & 1.00 & 0.98 & 0.03 & 0.92 & 1.00 & 0.87 & 0.08 & 0.77 & 0.97 \\
\rowcolor{red!15}
ICAN & 0.99 & 0.01 & 0.98 & 1.00 & 0.91 & 0.01 & 0.90 & 0.92 & 0.91 & 0.04 & 0.86 & 0.97 & 0.94 & 0.04 & 0.88 & 0.98 \\
\rowcolor{red!15}
IMULOG & 1.00 & 0.00 & 0.99 & 1.00 & 0.93 & 0.01 & 0.92 & 0.96 & 0.87 & 0.04 & 0.84 & 0.93 & 0.85 & 0.04 & 0.80 & 0.90 \\
\rowcolor{red!15}
LORASEN & 1.00 & 0.00 & 1.00 & 1.00 & 0.94 & 0.04 & 0.91 & 1.00 & 0.88 & 0.03 & 0.83 & 0.92 & 0.91 & 0.03 & 0.88 & 0.97 \\
\rowcolor{red!15}
VTRACK & 1.00 & 0.00 & 0.99 & 1.00 & 0.97 & 0.02 & 0.95 & 1.00 & 0.94 & 0.04 & 0.90 & 0.99 & 0.86 & 0.05 & 0.76 & 0.90 \\
\rowcolor{red!15}
\textbf{Hard Avg.} & \textbf{1.00} & \textbf{0.00} & \textbf{0.99} & \textbf{1.00} & \textbf{0.95} & \textbf{0.03} & \textbf{0.91} & \textbf{1.00} & \textbf{0.92} & \textbf{0.04} & \textbf{0.87} & \textbf{0.98} & \textbf{0.89} & \textbf{0.03} & \textbf{0.85} & \textbf{0.94} \\
\midrule
\textbf{Overall Avg.} & \textbf{1.00} & \textbf{0.01} & \textbf{0.97} & \textbf{1.00} & \textbf{0.98} & \textbf{0.03} & \textbf{0.91} & \textbf{1.00} & \textbf{0.97} & \textbf{0.04} & \textbf{0.87} & \textbf{1.00} & \textbf{0.95} & \textbf{0.05} & \textbf{0.85} & \textbf{1.00} \\
\bottomrule
\end{tabular}
}
\end{table*}

%% file: chapters/failure_taxonomy.tex
\subsection{Failure Analysis}
\label{sec:failure-analysis}

To understand why automatic generation still fails on some benchmark tasks, we analyze all generated circuits labeled by the human grader as not working.
The goal is diagnostic: when \systemName{} fails, does it fail because of missing helper circuitry, wrong values, interface mistakes, local topological errors, or inconsistent subsystem configuration?

In total, this subset contains \FailureTaxonomyFailedRuns{} failures out of \FailureTaxonomyEvaluatedRuns{} evaluated runs (\FailureTaxonomyFailureRatePct\%).
We analyze these failed runs using a multi-label taxonomy that separates electrical and structural problems into five categories:
\begin{itemize}
    \item \textbf{F1: Wrong supporting component presence/type}: a required helper component is missing, an unnecessary helper component is added, or the helper component kind is wrong.
    \item \textbf{F2: Wrong supporting component value}: the helper component is present and of the right general type, but its electrical value or rating is wrong.
    \item \textbf{F3: Wrong component configuration}: a configurable part is strapped or programmed incorrectly through pins or dedicated setting elements.
    \item \textbf{F4: Wrong interface connections}: required signal or power connections between subsystems are swapped, omitted, floating, or attached to the wrong endpoint.
    \item \textbf{F5: Wrong local topology}: the local circuit structure around a function block is wrong, such as a broken feedback arrangement, incorrect biasing structure, or reversed polarity.
\end{itemize}

\autoref{tab:human-failure-taxonomy} summarizes how often each category occurs in human-labeled non-working circuits.
Because a failing circuit may exhibit multiple error classes, the counts are multi-label and shares do not sum to 100\%.
The largest categories are \emph{wrong supporting component presence or type} (\FailureTaxonomySupportingComponentTypeCount{}, \FailureTaxonomySupportingComponentTypeSharePct\%) and \emph{wrong interface connections} (\FailureTaxonomyInterfaceCount{}, \FailureTaxonomyInterfaceSharePct\%).
\emph{Wrong supporting component value} appears in \FailureTaxonomySupportingComponentValueCount{} failures (\FailureTaxonomySupportingComponentValueSharePct\%), \emph{wrong local topology} in \FailureTaxonomyTopologyCount{} (\FailureTaxonomyTopologySharePct\%), and \emph{wrong component configuration} in \FailureTaxonomyConfigurationCount{} (\FailureTaxonomyConfigurationSharePct\%).
The dominant failures are therefore engineering-detail mistakes in support circuitry, values, interfaces, and local block realization rather than total architectural breakdowns.

\begin{table}[t]
\centering
\caption{Multi-label taxonomy of human-labeled non-working circuits. The taxonomy labels are assigned automatically, and shares are normalized by the number of non-working circuits.}
\label{tab:human-failure-taxonomy}
\begin{tabular}{lrr}
\toprule
Category & Count & Share \\
\midrule
F1: Wrong supporting component presence/type & \FailureTaxonomySupportingComponentTypeCount{} & \FailureTaxonomySupportingComponentTypeSharePct\% \\
F2: Wrong supporting component value & \FailureTaxonomySupportingComponentValueCount{} & \FailureTaxonomySupportingComponentValueSharePct\% \\
F3: Wrong component configuration & \FailureTaxonomyConfigurationCount{} & \FailureTaxonomyConfigurationSharePct\% \\
F4: Wrong interface connections & \FailureTaxonomyInterfaceCount{} & \FailureTaxonomyInterfaceSharePct\% \\
F5: Wrong local topology & \FailureTaxonomyTopologyCount{} & \FailureTaxonomyTopologySharePct\% \\
\bottomrule
\end{tabular}
\end{table}

\subsubsection{Failures by benchmark difficulty}
\begin{figure*}[t]
\centering
\begin{subfigure}[t]{0.49\textwidth}
\centering
\includegraphics[width=\textwidth]{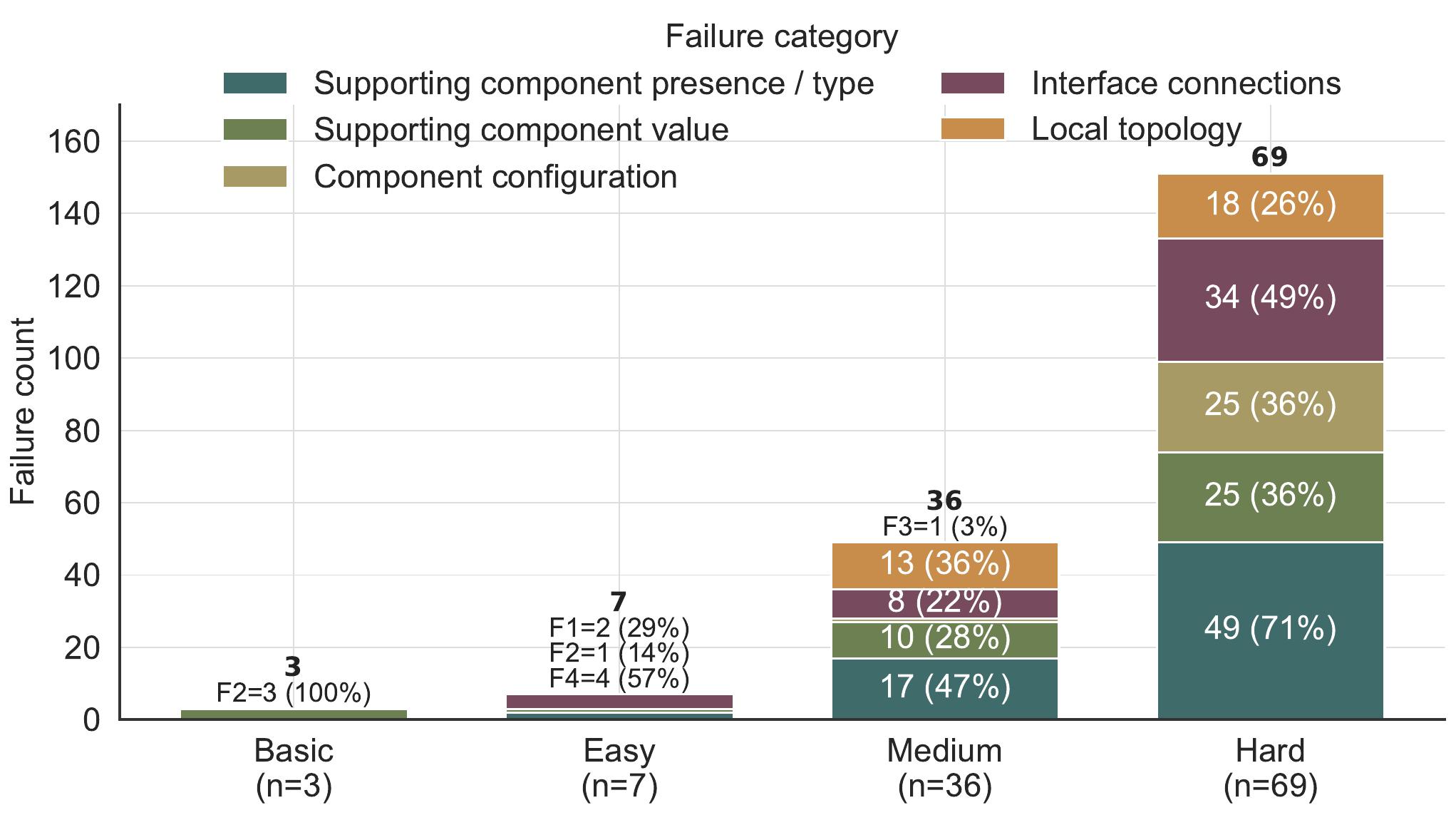}
\caption{Failure composition by difficulty.}
\label{fig:failure-taxonomy-heatmap}
\end{subfigure}\hfill
\begin{subfigure}[t]{0.49\textwidth}
\centering
\includegraphics[width=\textwidth]{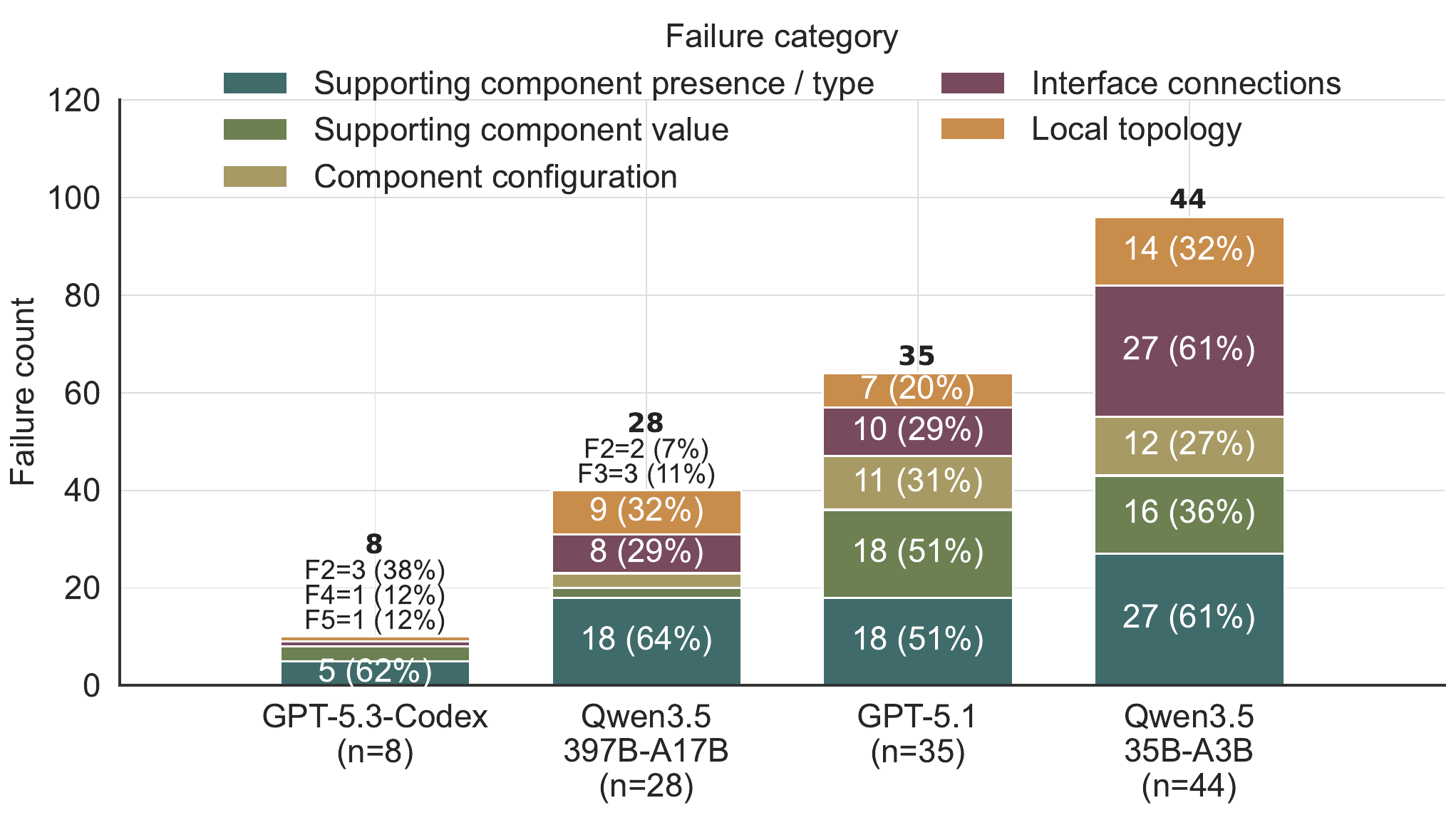}
\caption{Failure composition by model.}
\label{fig:failure-taxonomy-by-model}
\end{subfigure}
\caption{Visual summary of the failure analysis: (a) distribution of taxonomy categories across benchmark difficulty levels and (b) multi-label failure composition by model.}
\label{fig:failure-taxonomy-visual-summary}
\end{figure*}

\autoref{fig:failure-taxonomy-heatmap} shows that error categories are strongly concentrated by benchmark difficulty: harder prompts accumulate more errors of all types. This is not surprising, as harder tasks require composing several interacting subcircuits while preserving detailed interface constraints. As circuits grow and include more subsystems, design choices multiply, and mistakes in one subsystem can propagate and produce several simultaneous failure labels.

In easier tasks, interface-connection errors are less common since no component interconnection between subsystems is required. Wrong local topology concentrates in medium and hard tasks, while supporting-component categories span nearly the full benchmark range. The main effect of task difficulty is therefore a broader accumulation of interacting error modes as the design space grows.

This degradation is consistent with the electrical decisions that harder tasks introduce: more configurable components, more interacting subsystems, and more local support networks requiring simultaneous correct decisions. Larger designs also contain multiple power rails and mixed supply domains. The challenge in harder prompts is thus maintaining cross-subsystem electrical consistency rather than recovering the high-level architecture alone.

\subsubsection{Failures by model}
\label{sec:failure-taxonomy-by-model}

Beyond the aggregate taxonomy, \autoref{fig:failure-taxonomy-by-model} compares failure profiles across models. The overall human-labeled failure rates are \FailureTaxonomyModelGptFiveThreeCodexFailureRatePct\% for GPT-5.3-Codex, \FailureTaxonomyModelQwenThreeFiveLargeFailureRatePct\% for Qwen3.5-397B-A17B, \FailureTaxonomyModelGptFiveOneFailureRatePct\% for GPT-5.1, and \FailureTaxonomyModelQwenThreeFiveSmallFailureRatePct\% for Qwen3.5-35B-A3B, making GPT-5.3-Codex the most reliable model.

Stronger models reduce the total number of non-working runs substantially, but their remaining failures are still dominated by supporting-component, local-topology, and interface mistakes. Scaling improves reliability but does not eliminate the need for careful review of implementation details.

Once a circuit is labeled non-working, the most common annotations concentrate in supporting-component mistakes, with local-topology and interface mistakes also frequent. Many generated designs are close to workable drafts, but still fail in details such as swapped SPI or I\textsuperscript{2}C interface nets, incorrect regulator feedback resistor values, missing decoupling or pull-up components, and helper networks attached to the wrong pins.

The weaker Qwen3.5-35B-A3B model shows the broadest spread of failure types, covering substantial numbers of supporting-component, topology, and interface errors.
GPT-5.3-Codex fails much less often overall, but its remaining errors are still distributed across multiple categories. Supporting-component categories remain important even for stronger models: once more obvious interface mistakes are reduced, the remaining errors shift toward subtler helper-circuit and value-level details. Stronger models fail less often and their residual errors, such as component-values mistakes, are less likely to render the circuit completely inoperable.

\subsubsection{Failures by benchmark problem}
\begin{figure*}[t]
\centering
\includegraphics[width=\textwidth]{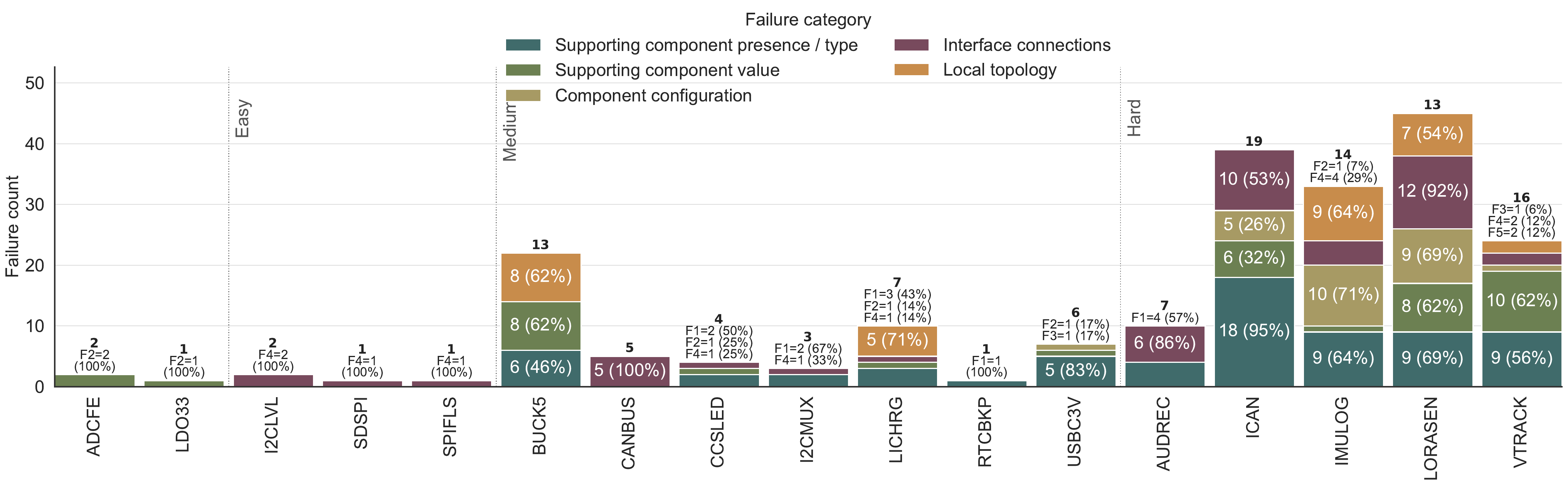}
\caption{Per-problem failure composition for human-labeled non-working circuits. Each bar shows multi-label failure counts for one benchmark prompt, normalized by the number of failed runs for that prompt.}
\Description{Bar chart showing normalized multi-label failure composition for each benchmark problem. Some problems are dominated by one failure type, while larger subsystem-integration prompts accumulate several failure categories at once.}
\label{fig:failure-taxonomy-by-problem}
\end{figure*}

\autoref{fig:failure-taxonomy-by-problem} breaks the failure composition down at the benchmark-problem level. Some benchmark problems are almost single-mode failures, while others, especially larger composed designs, accumulate several labels at once, indicating loss of consistency across multiple interacting constraints rather than one isolated mistake.

The clearest single-mode prompts are the small interface and passive tasks. ADCFE and LDO33 fail only in the supporting-component-value category, while I2CLVL, SDSPI, SPIFLS, and CANBUS are dominated by interface-connection errors. Some prompts are not broadly hard; they are narrow tests of whether the model can preserve one critical mapping rule.

Medium-difficulty prompts show a more heterogeneous profile. BUCK5's 13 failed runs are driven mainly by wrong local topology and wrong supporting-component values (8 each), with supporting-component presence/type also appearing in 6. LICHRG behaves similarly, with topology errors dominating 5 of 7 failures, while USBC3V is concentrated in supporting-component presence/type mistakes (5 of 6). Once power handling and regulation enter the benchmark, failures stop looking like clean protocol swaps and start looking like datasheet-level implementation errors inside otherwise plausible blocks.

Hard prompts split into two behaviors. AUDREC is relatively concentrated, with interface errors in 6 of 7 failures and supporting-component presence/type errors in 4. By contrast, IMULOG and LORASEN are genuinely multi-label failures. IMULOG spreads across supporting-component presence/type (9/14), component configuration (10/14), local topology (9/14), and interface mistakes (4/14), while LORASEN activates all five categories at high rates. These are the clearest examples of compounding design breakdown: the model is failing to keep power, configuration, support circuitry, and interface behavior mutually consistent across the full design.

VTRACK's errors are broad but not interface-heavy: supporting-component presence/type (9/16) and supporting-component value (10/16) dominate, while topology and interface labels are comparatively rare. This suggests that for some large prompts the model can preserve the overall architecture and many major interconnects, yet miss the benchmark because surrounding resistors, capacitors, and helper networks lack sufficient electrical precision. Top-level component recovery alone is too weak a proxy for whether a generated embedded design is operable.

\subsubsection{Exploratory Failure Correlates}
\begin{figure*}[t]
\centering
\begin{subfigure}[t]{0.49\textwidth}
\vspace{0pt}
\centering
\begin{minipage}[t][5.45cm][t]{\textwidth}
\centering
\includegraphics[width=\textwidth]{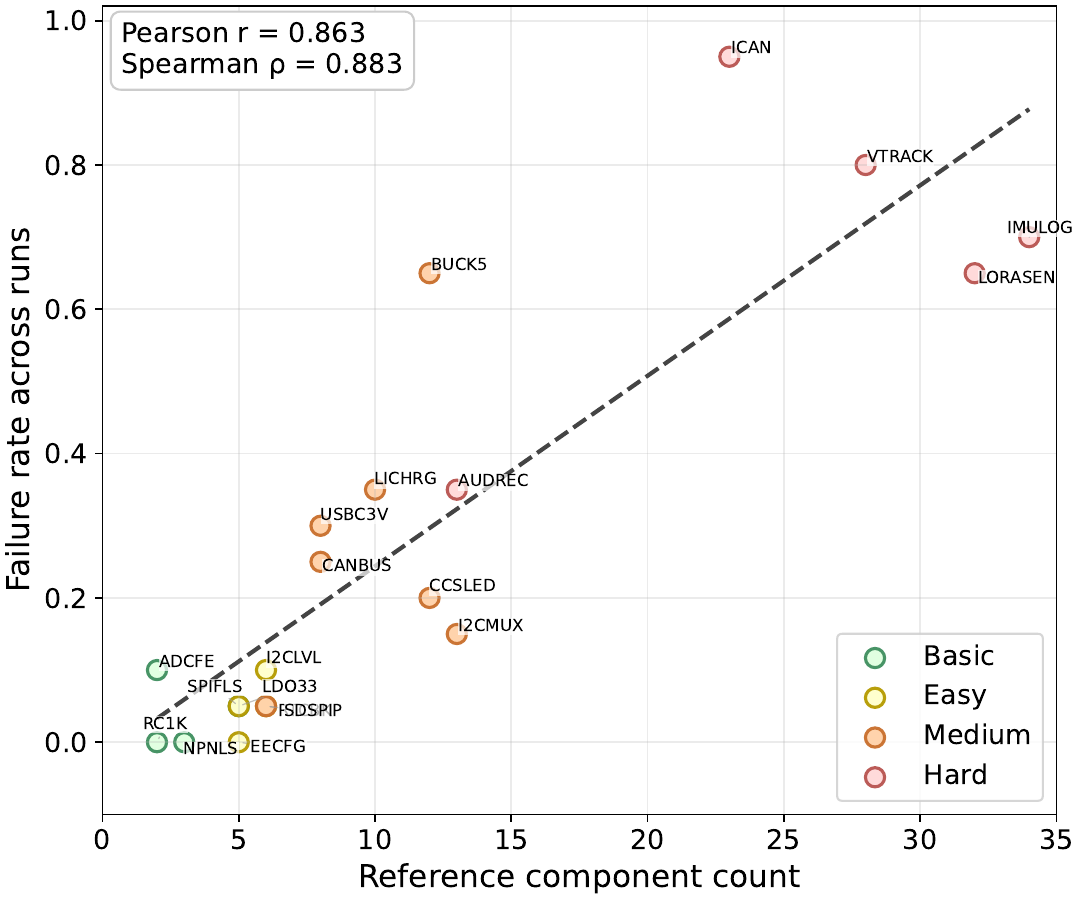}
\end{minipage}
\caption{Prompt-level relationship between reference component count and failure rate. Larger benchmark circuits fail more often.}
\Description{Scatter plot relating the number of components in each reference circuit to the corresponding prompt failure rate. Failure rate increases with circuit size.}
\label{fig:failure-components-vs-rate}
\end{subfigure}\hfill
\begin{subfigure}[t]{0.49\textwidth}
\vspace{0pt}
\centering
\begin{minipage}[t][5.45cm][t]{\textwidth}
\centering
\includegraphics[width=\textwidth]{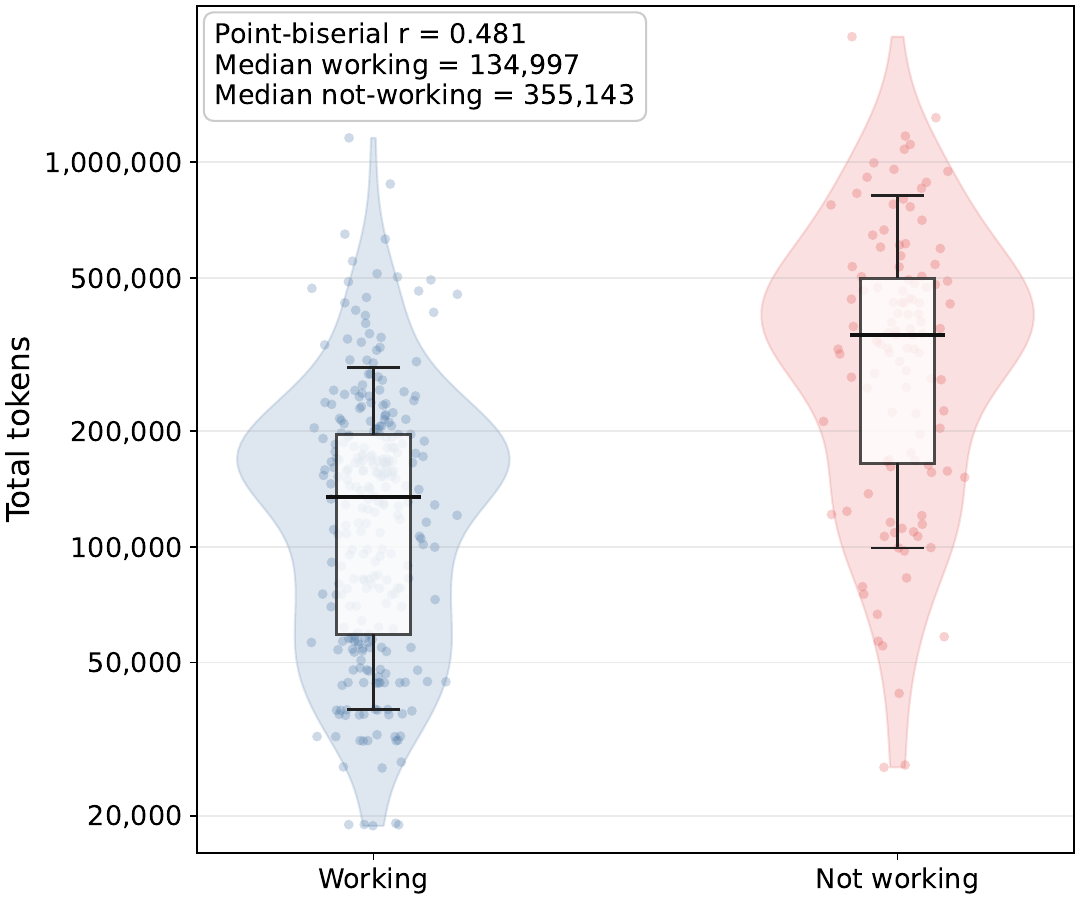}
\end{minipage}
\caption{Run-level token distributions for working and non-working generations. Violin width reflects density, the box marks the interquartile range, and the center line marks the median. Non-working runs consume substantially more cumulative token budget.}
\Description{Violin plot comparing cumulative token usage for working and non-working runs. Non-working runs have visibly higher token consumption and a broader distribution.}
\label{fig:failure-token-burn}
\end{subfigure}
\caption{Exploratory correlates of failure. Left: prompt-level structural complexity. Right: cumulative generation effort at run level. Both views are consistent with failures becoming more likely when the system must sustain longer repair trajectories or synthesize larger circuits. The analysis remains correlational and does not isolate the effects of context length, prompting hierarchy, or subsystem decomposition.}
\label{fig:failure-complexity-proxies}
\end{figure*}

The taxonomy and per-problem discussion above suggest that failures become more likely when the system must sustain longer design trajectories and coordinate more electrical detail within a single schematic.
\autoref{fig:failure-complexity-proxies} provides two quantitative views that are consistent with that interpretation.

First, \autoref{fig:failure-components-vs-rate} shows that prompt-level failure rate rises strongly with benchmark circuit size.
Across the \FailureCorrelationPromptCount{} prompts with implemented reference circuits, failure rate correlates strongly with reference component count (Pearson $r = \FailureCorrelationComponentPearsonR{}$, Spearman $\rho = \FailureCorrelationComponentSpearmanRho{}$).
This matches the qualitative analysis above: larger circuits require the model to keep more components, support networks, and subsystem interfaces mutually consistent, increasing opportunities for local mistakes to accumulate.

Second, \autoref{fig:failure-token-burn} shows a moderate positive association between cumulative token usage and human-labeled failure.
Across all \FailureCorrelationEvaluatedRuns{} evaluated runs, the point-biserial correlation between total token usage and human-labeled failure is \FailureCorrelationTokenPointBiserialR{}, and median total token usage rises from \FailureCorrelationWorkingMedianTokens{} tokens for working runs to \FailureCorrelationNonWorkingMedianTokens{} for non-working runs.
This gap is consistent with harder prompts requiring more repair attempts and longer generation trajectories.
However, these totals are cumulative over the entire run, not the number of tokens seen at once in a single step, so they should not be interpreted as a direct measure of context length or context-window saturation.

These plots do not establish causality, but they strengthen the interpretation that failures are systematically associated with higher generation effort and structurally larger circuit tasks.

Overall, the failure analysis supports the same practical conclusion as the automatic evaluation: \systemName{} already produces useful first drafts, but the remaining errors are concentrated in exactly the kinds of detailed electrical decisions that still require expert review.

%% file: chapters/discussion.tex
\section{Discussion}

\subsection{Implications for Hardware Design Workflows}

\systemName{} is best understood as a grounded schematic-generation system for early hardware-development workflows.
Rather than replacing established EDA tools, it lowers the effort required to move from a natural-language requirement to an initial editable schematic.
This makes the system particularly relevant for rapid prototyping, early ideation, and interdisciplinary settings in which users understand the intended device behavior but may still rely on expert or hardware-literate review to turn that intent into a correct schematic structure.
In this sense, the system has two distinct audiences: interdisciplinary users can use it to express and explore device ideas at the schematic level, while hardware-literate users and experts remain responsible for inspecting, correcting, and accepting the resulting designs.

For a substantial portion of the benchmark, especially simpler and moderately composed designs, the system can already produce structurally correct or near-correct schematics that are directly editable in KiCad.
With \AutomaticResultsBestModelName{}, the system reaches an overall pass@1 of \AutomaticResultsBestOverallPassAtOne{} and pass@5 of \AutomaticResultsBestOverallPassAtFive{}, indicating that automatic generation is already strong enough to produce reviewable first drafts for many embedded design tasks.
At the same time, the results and failure analysis show that these drafts are not yet reliable enough to be used without expert inspection.
The remaining errors are concentrated in supporting circuitry, values, interface details, and local topology, which are precisely the kinds of mistakes that determine whether a schematic is practically usable.
In practical terms, many of the generated circuits already reflect generic component-integration requirements drawn from datasheets, such as expected support circuitry, standard interface wiring, and nominal power-domain structure.
However, they are not yet reliably tuned to the exact application context of a given use case, where task-specific tradeoffs, parameter choices, and subsystem interactions still require expert adaptation.

The strongest practical interpretation is therefore not fully autonomous design, but design acceleration.
\systemName{} can reduce the effort required to reach a reviewable starting point, expose the right subsystem decomposition, and externalize many of the routine wiring decisions that would otherwise be assembled manually from datasheets and reference designs.
The strong pass@5 results further suggest that the system is most useful in workflows where users inspect, compare, and refine multiple candidates rather than relying on a single one-shot output.
Taken together, these results show that the overall approach already works in principle: \systemName{} can generate schematics that form a useful starting point for further development and iteration, even though performance still varies substantially with task complexity and model choice.
The current evidence therefore supports a precise practical claim: \systemName{} can already generate reviewable first-draft schematics that are useful for further development, but it does not yet provide a level of reliability that would justify replacing expert schematic engineering judgment.

\subsection{Role of Interactive Tooling}

The interactive web interface is an important systems feature because it connects the automatic generation system to how hardware design is actually carried out in practice.
Hardware design is rarely solved in a single generation pass; users inspect outputs, compare alternatives, cross-check component choices, and incrementally revise the design.
By supporting browser-based interaction, project synchronization, and continued work on existing KiCad projects, the system fits more naturally into real design practice than a one-shot text-to-schematic model alone.
In particular, the web UI allows hardware-literate users and experts to directly chat with the model, iteratively request changes, and improve the design while remaining inside a KiCad-centered workflow.
The web-based workflow therefore extends the generator toward practical use by keeping the human in the loop and making the system usable within KiCad-centered prototyping workflows.

\subsection{Limitations}

Several limitations remain.
First, the system targets schematic generation rather than full PCB realization. It does not automate placement, routing, manufacturing preparation, or fabrication-specific optimization, and it does not reason directly about layout-dependent effects that often determine whether a design remains robust beyond the schematic stage.
Second, the validation pipeline establishes executability, rule-level consistency, and schematic-level plausibility, but it is not a guarantee of electrical correctness. The semantic validator can sometimes infer constraints such as signal direction, support circuitry, or interface handling from prompts, datasheets, and circuit context, but these checks remain indirect. Effects such as signal integrity, EMI, thermal behavior, tolerance stacking, startup transients, and other layout- or operating-condition-sensitive issues remain outside the scope of the current system.
Third, the benchmark should be interpreted as a conservative, domain-specific evaluation rather than a complete measure of hardware-design competence. It focuses on embedded, IoT, and wearable-style schematic tasks, and the reference circuits are datasheet-grounded exemplar solutions rather than an exhaustive set of all acceptable realizations. Although the benchmark already models some flexibility through optional parts and flexible pin assignments, valid alternative circuits may still be underrewarded, especially when they differ mainly in support circuitry, parameter choices, or interface realization.
Finally, datasheet grounding and component-library access improve realism, but they do not eliminate ambiguity or ensure use-case-specific correctness. The quality of generated outputs still depends on the availability and correctness of component metadata, datasheets, and KiCad library representations, and also on how completely the prompt captures the intended operating conditions and design tradeoffs. A design may therefore follow generic component-integration guidance while still being insufficiently tuned to the exact application context. We also do not yet measure downstream efficiency gains in realistic design workflows.

\subsection{Future Work}

There are three especially important directions for future work.
First, a natural next step is to investigate how model adaptation on schematic-generation tasks, component-integration patterns, and datasheet-grounded circuit knowledge affects reliability on the detailed electrical decisions that still cause many failures.
Second, the benchmark should be expanded to cover a broader range of circuit classes and larger multi-subsystem designs, enabling more extensive evaluation of both automatic generation quality and valid design variation.
Third, the interactive workflow should be studied directly with a focus on expert efficiency gains, for example through user studies that measure whether natural-language prompting, schematic inspection, manual editing, and project synchronization reduce the time and effort required for hardware-literate users to reach a correct design.

%% file: appendix/benchmark_inventory.tex
\section{Benchmark Inventory}
\label{app:benchmark-inventory}
\input{tables/appendix_benchmark_inventory.tex}

%% file: tables/appendix_benchmark_inventory.tex
\subsection{Basic}

\setlength{\tabcolsep}{5pt}
\renewcommand{\arraystretch}{1.15}
\noindent
\begin{tabularx}{\textwidth}{@{}p{0.2\textwidth}X@{}}
\toprule
\textbf{ADCFE} & \textbf{ADC front-end conditioning} \\
\midrule
\textbf{Prompt} & Create a resistor-divider circuit that maps a 0-5 V analog sensor signal to a 0-3.3 V ADC input. When SENSOR\_IN is 5 V, ADC\_IN must be 3.3 V. Use exactly two resistors: 17 kOhm and 33 kOhm. \\
\textbf{Components} & 2 \\
\textbf{Pins} & 4 \\
\textbf{Connected pins} & 4 \\
\textbf{Required parts} & 17k, 33k \\
\textbf{Exposed nets} & GND, SENSOR\_IN, ADC\_IN \\
\bottomrule
\end{tabularx}
\vspace{0.8em}

\noindent
\begin{tabularx}{\textwidth}{@{}p{0.2\textwidth}X@{}}
\toprule
\textbf{LDO33} & \textbf{5V-to-3.3V LDO regulator} \\
\midrule
\textbf{Prompt} & Design a 5 V to 3.3 V linear regulator stage using AMS1117-3.3. \\
\textbf{Components} & 5 \\
\textbf{Pins} & 11 \\
\textbf{Connected pins} & 11 \\
\textbf{Required parts} & AMS1117-3.3 \\
\textbf{Exposed nets} & +5V, GND, +3V3 \\
\bottomrule
\end{tabularx}
\vspace{0.8em}

\noindent
\begin{tabularx}{\textwidth}{@{}p{0.2\textwidth}X@{}}
\toprule
\textbf{NPNLS} & \textbf{NPN open-collector level shifter} \\
\midrule
\textbf{Prompt} & Design a simple NPN transistor inverter/level shifter that accepts a 3.3 V logic input and provides a 5 V open-collector output with a pull-up resistor. Use 10 kOhm resistors. \\
\textbf{Components} & 3 \\
\textbf{Pins} & 7 \\
\textbf{Connected pins} & 7 \\
\textbf{Required parts} & PN2222A \\
\textbf{Exposed nets} & +5V, GND, IN, OUT \\
\bottomrule
\end{tabularx}
\vspace{0.8em}

\noindent
\begin{tabularx}{\textwidth}{@{}p{0.2\textwidth}X@{}}
\toprule
\textbf{RC1K} & \textbf{1 kHz RC low-pass} \\
\midrule
\textbf{Prompt} & Design a first-order passive RC low-pass filter with cutoff frequency f\_c = 1 kHz using a 10 kOhm resistor. Assume the signal source has negligible output impedance and the load on OUT is at least 100 kOhm. \\
\textbf{Components} & 2 \\
\textbf{Pins} & 4 \\
\textbf{Connected pins} & 4 \\
\textbf{Required parts} & R, C \\
\textbf{Exposed nets} & IN, GND, OUT \\
\bottomrule
\end{tabularx}
\vspace{0.8em}

\subsection{Easy}

\setlength{\tabcolsep}{5pt}
\renewcommand{\arraystretch}{1.15}
\noindent
\begin{tabularx}{\textwidth}{@{}p{0.2\textwidth}X@{}}
\toprule
\textbf{EECFG} & \textbf{I2C EEPROM configuration} \\
\midrule
\textbf{Prompt} & Add an I2C EEPROM to a 3.3 V microcontroller for configuration storage. \\
\textbf{Components} & 5 \\
\textbf{Pins} & 44 \\
\textbf{Connected pins} & 19 \\
\textbf{Required parts} & 24LC256, Arduino\_Nano\_ESP32 \\
\textbf{Exposed nets} & +3V3, GND, SDA, SCL \\
\bottomrule
\end{tabularx}
\vspace{0.8em}

\noindent
\begin{tabularx}{\textwidth}{@{}p{0.2\textwidth}X@{}}
\toprule
\textbf{I2CLVL} & \textbf{Bidirectional I2C level shifter} \\
\midrule
\textbf{Prompt} & Design a bidirectional I2C level shifter between a 3.3 V microcontroller bus and a 5 V I2C bus. Include pull-up resistors on both sides. \\
\textbf{Components} & 6 \\
\textbf{Pins} & 14 \\
\textbf{Connected pins} & 14 \\
\textbf{Required parts} & BSS138, R \\
\textbf{Exposed nets} & +3V3, +5V, SDA\_3V3, SCL\_3V3, SDA\_5V, SCL\_5V \\
\bottomrule
\end{tabularx}
\vspace{0.8em}

\noindent
\begin{tabularx}{\textwidth}{@{}p{0.2\textwidth}X@{}}
\toprule
\textbf{SDSPI} & \textbf{SD card SPI interface} \\
\midrule
\textbf{Prompt} & Connect an SD card socket to a 3.3 V microcontroller using SPI communication. Do not use the card-detect pin; leave it floating. Connect the socket shield to GND. \\
\textbf{Components} & 6 \\
\textbf{Pins} & 48 \\
\textbf{Connected pins} & 22 \\
\textbf{Required parts} & Micro\_SD\_Card\_Det1, Arduino\_Nano\_ESP32 \\
\textbf{Exposed nets} & +3V3, GND, SCK, MOSI, MISO, CS \\
\bottomrule
\end{tabularx}
\vspace{0.8em}

\noindent
\begin{tabularx}{\textwidth}{@{}p{0.2\textwidth}X@{}}
\toprule
\textbf{SPIFLS} & \textbf{SPI flash memory interface} \\
\midrule
\textbf{Prompt} & Add a 3.3 V SPI flash memory to a microcontroller. Include the standard SPI connections. Add any pull-ups needed for normal single-bit SPI operation. \\
\textbf{Components} & 5 \\
\textbf{Pins} & 44 \\
\textbf{Connected pins} & 21 \\
\textbf{Required parts} & W25Q16JVSS, Arduino\_Nano\_ESP32 \\
\textbf{Exposed nets} & +3V3, GND, SCK, MOSI, MISO, CS \\
\bottomrule
\end{tabularx}
\vspace{0.8em}

\subsection{Medium}

\setlength{\tabcolsep}{5pt}
\renewcommand{\arraystretch}{1.15}
\noindent
\begin{tabularx}{\textwidth}{@{}p{0.2\textwidth}X@{}}
\toprule
\textbf{BUCK5} & \textbf{12V-to-5V buck converter} \\
\midrule
\textbf{Prompt} & Design a 12 V to 5 V buck regulator stage using the TPS5430, including the supporting components needed for normal operation. \\
\textbf{Components} & 12 \\
\textbf{Pins} & 31 \\
\textbf{Connected pins} & 29 \\
\textbf{Required parts} & TPS5430DDA, B340, L\_Iron \\
\textbf{Exposed nets} & +12V, GND, +5V \\
\bottomrule
\end{tabularx}
\vspace{0.8em}

\noindent
\begin{tabularx}{\textwidth}{@{}p{0.2\textwidth}X@{}}
\toprule
\textbf{CANBUS} & \textbf{CAN bus MCU interface} \\
\midrule
\textbf{Prompt} & Communicate between two microcontrollers using a CAN bus interface. Use SN65HVD232 transceivers powered from +3V3, keep each Arduino Nano ESP32 powered from +5V, connect each MCU +3V3 pin, and allow transceiver pins 5 and 8 to be left unconnected or tied to GND. Place a 100 nF decoupling capacitor from +3V3 to GND at each transceiver, and include a 120 \ensuremath{\Omega} termination resistor at each end of the CAN bus. \\
\textbf{Components} & 8 \\
\textbf{Pins} & 84 \\
\textbf{Connected pins} & 32 \\
\textbf{Required parts} & SN65HVD232, Arduino\_Nano\_ESP32 \\
\textbf{Exposed nets} & +5V, +3V3, GND \\
\bottomrule
\end{tabularx}
\vspace{0.8em}

\noindent
\begin{tabularx}{\textwidth}{@{}p{0.2\textwidth}X@{}}
\toprule
\textbf{CCSLED} & \textbf{CCS811 air-quality indicator} \\
\midrule
\textbf{Prompt} & Create an air-quality indicator using a CCS811 gas sensor and a microcontroller. Display the air-quality state using three LEDs: green, yellow, and red. Drive the green, yellow, and red LEDs from Arduino Nano ESP32 pins D2, D3, and D4, respectively. Connect the wake pin to the MCU. Do not connect the interrupt pin, as this functionality is not needed. Also do not connect the reset pin to the MCU, as this functionality is not needed. Do not add any reset pull-up resistor or other reset circuitry. \\
\textbf{Components} & 12 \\
\textbf{Pins} & 61 \\
\textbf{Connected pins} & 37 \\
\textbf{Required parts} & CCS811, Arduino\_Nano\_ESP32, LED \\
\textbf{Exposed nets} & +3V3, GND, SDA, SCL \\
\bottomrule
\end{tabularx}
\vspace{0.8em}

\noindent
\begin{tabularx}{\textwidth}{@{}p{0.2\textwidth}X@{}}
\toprule
\textbf{I2CMUX} & \textbf{I2C mux dual same-address sensors} \\
\midrule
\textbf{Prompt} & Design a microcontroller circuit that reads two identical I2C sensors with the same fixed address using an I2C multiplexer. Include pull-up resistors, power, and separate downstream bus branches. Use the Arduino Nano ESP32 as the microcontroller. Assume the user powers this circuit using a USB-C connection plugged directly into the Arduino Nano ESP32. Do not add reset pull-up resistors or other reset circuitry for the I2C multiplexer. \\
\textbf{Components} & 13 \\
\textbf{Pins} & 81 \\
\textbf{Connected pins} & 44 \\
\textbf{Required parts} & Arduino\_Nano\_ESP32, TCA9548AMRGER, MS5837-xxBA, MS5837-xxBA \\
\textbf{Exposed nets} & +3V3, GND, SDA, SCL, SDA\_CH0, SCL\_CH0, SDA\_CH1, SCL\_CH1 \\
\bottomrule
\end{tabularx}
\vspace{0.8em}

\noindent
\begin{tabularx}{\textwidth}{@{}p{0.2\textwidth}X@{}}
\toprule
\textbf{LICHRG} & \textbf{USB-C Li-Ion charger} \\
\midrule
\textbf{Prompt} & Create a charging circuit for a lithium-ion battery powered via USB-C (5 V only, include CC pull-down resistors). Include the battery as well as an LED to indicate charging status (LED on when charging, off when not charging). Do not add an input resistor between the USB-C 5 V rail and the charger input. \\
\textbf{Components} & 10 \\
\textbf{Pins} & 32 \\
\textbf{Connected pins} & 31 \\
\textbf{Required parts} & USB\_C\_Receptacle\_PowerOnly\_6P, TP4056-42-ESOP8, Battery\_Cell, LED \\
\textbf{Exposed nets} & -- \\
\bottomrule
\end{tabularx}
\vspace{0.8em}

\noindent
\begin{tabularx}{\textwidth}{@{}p{0.2\textwidth}X@{}}
\toprule
\textbf{RTCBKP} & \textbf{RTC backup-domain interface} \\
\midrule
\textbf{Prompt} & Add an I2C real-time clock to a microcontroller. Include the backup battery but do not add a charging circuit for it. Assume the RTC is powered by the backup battery when the main power is off. Include the RTC connection and I2C pull-up resistors. \\
\textbf{Components} & 6 \\
\textbf{Pins} & 54 \\
\textbf{Connected pins} & 28 \\
\textbf{Required parts} & Arduino\_Nano\_ESP32, DS3231M, Battery\_Cell \\
\textbf{Exposed nets} & +3V3, GND \\
\bottomrule
\end{tabularx}
\vspace{0.8em}

\noindent
\begin{tabularx}{\textwidth}{@{}p{0.2\textwidth}X@{}}
\toprule
\textbf{USBC3V} & \textbf{USB-C to 3.3V supply} \\
\midrule
\textbf{Prompt} & Use a USB-C connector (5 V only, no USB-PD) with CC pull-down resistors to output 3.3 V for a microcontroller project. \\
\textbf{Components} & 8 \\
\textbf{Pins} & 22 \\
\textbf{Connected pins} & 22 \\
\textbf{Required parts} & USB\_C\_Receptacle\_PowerOnly\_6P, AMS1117-3.3 \\
\textbf{Exposed nets} & +3V3, GND \\
\bottomrule
\end{tabularx}
\vspace{0.8em}

\subsection{Hard}

\setlength{\tabcolsep}{5pt}
\renewcommand{\arraystretch}{1.15}
\noindent
\begin{tabularx}{\textwidth}{@{}p{0.2\textwidth}X@{}}
\toprule
\textbf{AUDREC} & \textbf{Portable audio recorder} \\
\midrule
\textbf{Prompt} & Design a portable audio recorder using an ESP32 microcontroller. The design should feature two I2S microphones and a microSD card for storage. Wire one microphone as the left channel and the other as the right channel. Attach the SD card to the microcontroller using SPI and set up the card-detect line. Include pull-up resistors on the SD card SPI lines. \\
\textbf{Components} & 13 \\
\textbf{Pins} & 70 \\
\textbf{Connected pins} & 50 \\
\textbf{Required parts} & Arduino\_Nano\_ESP32, ICS-43434, Micro\_SD\_Card\_Det1 \\
\textbf{Exposed nets} & +5V, GND \\
\bottomrule
\end{tabularx}
\vspace{0.8em}

\noindent
\begin{tabularx}{\textwidth}{@{}p{0.2\textwidth}X@{}}
\toprule
\textbf{ICAN} & \textbf{USB CAN sensor node} \\
\midrule
\textbf{Prompt} & Design an industrial sensor node using a microcontroller, a CAN transceiver, and an environmental sensor. Use the ESP32-S2-WROVER as the microcontroller module, the TCAN332 as the CAN transceiver, and the BME280 as the environmental sensor. Power the circuit from a USB-C connector wired for native USB flashing of the ESP32-S2. Include explicit BOOT and RESET buttons for the MCU, both with pull-up resistors to 3.3 V. Add a linear regulator from USB VBUS to 3.3 V. Expose CANH/CANL on a separate 2-pin connector. CANH should be pin 1, and CANL should be pin 2. Do not include terminators for the CAN bus. \\
\textbf{Components} & 23 \\
\textbf{Pins} & 115 \\
\textbf{Connected pins} & 82 \\
\textbf{Required parts} & ESP32-S2-WROVER, TCAN332, BME280, TLV75733PDBV, USB\_C\_Receptacle\_USB2.0\_14P, Conn\_01x02, SW\_Push \\
\textbf{Exposed nets} & -- \\
\bottomrule
\end{tabularx}
\vspace{0.8em}

\noindent
\begin{tabularx}{\textwidth}{@{}p{0.2\textwidth}X@{}}
\toprule
\textbf{IMULOG} & \textbf{BLE IMU data logger} \\
\midrule
\textbf{Prompt} & Create a BLE IMU data logger powered from a Li-Ion battery: ESP32-C3 + BMI160 + microSD storage, with USB-C charging. Do not include a thermistor, but include three charger status LEDs: one for power-good, one for charge status 1, and one for charge status 2. Enable charging by default when USB power is present. Disable the safety timer. Set the charge current as well as the maximum input current to 500 mA. Stop charging when the current drops below 100 mA. Do not include a crystal or a button for the ESP32. Only include the parts for the ESP32 that are really necessary for it to run. Connect the SD card via SPI. Add a pull-up resistor on the SD card chip-select line so the card stays deselected during reset. Connect the IMU via I2C. Use the default I2C address for the IMU. For the SD card, do not connect the card-detect pin. \\
\textbf{Components} & 34 \\
\textbf{Pins} & 132 \\
\textbf{Connected pins} & 113 \\
\textbf{Required parts} & ESP32-C3-WROOM-02, BMI160, Micro\_SD\_Card\_Det1, MCP73871-2AA, TLV75733PDBV, Battery\_Cell, USB\_C\_Receptacle\_PowerOnly\_6P, LED \\
\textbf{Exposed nets} & -- \\
\bottomrule
\end{tabularx}
\vspace{0.8em}

\noindent
\begin{tabularx}{\textwidth}{@{}p{0.2\textwidth}X@{}}
\toprule
\textbf{LORASEN} & \textbf{LoRa sensor node} \\
\midrule
\textbf{Prompt} & Design a long-range sensor node with LoRa, an environmental sensor, and a Li-Ion battery. Use the RAK811-HF-EU868 as the LoRa module and the BME280 as the environmental sensor. Include a battery charger circuit for the Li-Ion battery with USB-C charging. Do not include a thermistor, but include three charger status LEDs: one for power good, one for charging in progress, and one for charge complete / standby. Disable the THERM pin of the charger. Do not measure the battery temperature. Enable charging by default when USB power is present. Disable the safety timer. Set the charge current as well as the maximum input current to 500 mA. Stop charging when the current drops below 100 mA. Connect the sensor to the LoRa module via I2C using the default I2C pins for the sensor and the LoRa module. Use the default I2C address for the sensor. \\
\textbf{Components} & 32 \\
\textbf{Pins} & 128 \\
\textbf{Connected pins} & 108 \\
\textbf{Required parts} & RAK811-HF-EU868, BME280, MCP73871-2AA, TLV75733PDBV, Battery\_Cell, USB\_C\_Receptacle\_PowerOnly\_6P, Antenna \\
\textbf{Exposed nets} & -- \\
\bottomrule
\end{tabularx}
\vspace{0.8em}

\noindent
\begin{tabularx}{\textwidth}{@{}p{0.2\textwidth}X@{}}
\toprule
\textbf{VTRACK} & \textbf{Vehicle tracking node} \\
\midrule
\textbf{Prompt} & Design a vehicle tracking node powered from a nominal 12 V input supply. Use the ESP32-C3-WROOM-02 as the microcontroller module, the LEA-M8T as the GNSS receiver module for position input, and the RAK811-HF-EU868 as the LoRa module for uplink reporting. Generate a 3.3 V rail from the 12 V input using an LMR33630ADDA buck regulator and power the MCU, GNSS module, and LoRa module from that 3.3 V rail. Add an ESP32 enable network with a pull-up resistor from EN to 3.3 V and a capacitor from EN to GND. Connect the GNSS receiver to the MCU over one UART and connect the RAK811 to the MCU over a second UART. Do not add any antenna matching, bias, or protection circuitry. Connect a simple antenna component directly to the GNSS RF pin and another antenna component directly to the LoRa RF pin. \\
\textbf{Components} & 28 \\
\textbf{Pins} & 136 \\
\textbf{Connected pins} & 92 \\
\textbf{Required parts} & ESP32-C3-WROOM-02, LMR33630ADDA, LEA-M8T, RAK811-HF-EU868, Antenna \\
\textbf{Exposed nets} & +12V, +3V3, GND, GNSS\_TX, GNSS\_RX, LORA\_TX, LORA\_RX \\
\bottomrule
\end{tabularx}
\vspace{0.8em}